\documentclass[acmsmall]{acmart} 

\AtBeginDocument{%
  }

\setcopyright{acmlicensed}
\copyrightyear{2018}
\acmYear{2018}
\acmDOI{XXXXXXX.XXXXXXX}

\acmJournal{JDS}
\acmVolume{37}
\acmNumber{4}
\acmArticle{111}
\acmMonth{8}

\usepackage{hyperref}
\usepackage{xcolor}

\usepackage{graphicx}
\usepackage{caption}
\usepackage{subcaption}
\usepackage{amsfonts}
\usepackage{amsmath}
\usepackage{amsthm}
\usepackage{url}
\usepackage{tabularx}
\usepackage{enumitem}
\usepackage{array}
\usepackage{booktabs}
\usepackage{float}
\usepackage{multicol}
\usepackage{multirow}
\usepackage{ragged2e}
\usepackage{bbding}
\usepackage{pifont}
\usepackage{wasysym}

\usepackage{forest}
\usepackage{caption} % 引入caption包以更好地控制标题
\usepackage{subcaption} % 引入subcaption包以支持子图标题

\newcommand{\saizhuoDrafting}{}
\ifdefined\saizhuoDrafting
\newcommand{\todo}[1]{{\color{blue}TODO: {#1}}}
\newcommand{\saizhuo}[1]{{\color{red}Saizhuo: {#1}}}
\else
\newcommand{\todo}[1]{}
\newcommand{\saizhuo}[1]{}
\fi

\ifdefined\qiComments
\newcommand{\qi}[1]{{\color{green}Qi: {#1}}}
\else
\newcommand{\qi}[1]{}
\fi

\newcommand{\saizhuoFeedback}{}
\ifdefined\saizhuoFeedback
\newcommand{\feedback}[1]{{\color{orange}Reply: #1}}
\else
\newcommand{\feedback}[1]{}
\fi

\newcolumntype{P}[1]{>{\centering \arraybackslash}p{#1}}
\newcolumntype{L}[1]{>{\raggedright \arraybackslash}p{#1}}
\newcolumntype{C}{>{\centering \arraybackslash}X}
\newcolumntype{R}{>{\raggedright \arraybackslash}X}
\newcolumntype{M}[1]{>{\centering\arraybackslash}m{#1}}

\AtBeginDocument{\hypersetup{pdfborder={0 0 1}}}
\usepackage{setspace}
\usepackage{tikz}
\usepackage{forest}

\usepackage{caption} % 引入caption包以更好地控制标题
\usepackage{subcaption} % 引入subcaption包以支持子图标题
\usetikzlibrary{shapes.geometric, arrows, positioning, calc, shadows}

\begin{document}
\title{From Deep Learning to LLMs: A survey of AI in Quantitative Investment}
% Í

\author{Bokai Cao}
\email{mabkcao@connect.hkust-gz.edu.cn}
\affiliation{%
  \institution{The Hong Kong University of Science and Technology (Guangzhou)}
  \country{China}
}
\affiliation{%
  \institution{IDEA Research, International Digital Economy Academy}
  \country{China}
}

\author{Saizhuo Wang}
\email{swangeh@connect.ust.hk}
\affiliation{%
  \institution{The Hong Kong University of Science and Technology}
  \country{Hong Kong}
}
\affiliation{%
  \institution{IDEA Research, International Digital Economy Academy}
  \country{China}
}

\author{Xinyi Lin}
\email{xlin652@connect.hkust-gz.edu.cn}
\author{Xiaojun Wu}
\email{xwu647@connect.hkust-gz.edu.cn}
\author{Haohan Zhang}
\email{hzhang760@connect.hkust-gz.edu.cn}
\affiliation{%
  \institution{The Hong Kong University of Science and Technology (Guangzhou)}
  \country{China}
}
\affiliation{%
  \institution{IDEA Research, International Digital Economy Academy}
  \country{China}
}

\author{Lionel M. Ni}
\affiliation{%
  \institution{The Hong Kong University of Science and Technology (Guangzhou)}
  \country{China}
}
\email{ni@ust.hk}

\author{Jian Guo}
\authornote{Corresponding Author.}
\affiliation{%
  \institution{IDEA Research, International Digital Economy
Academy}
  \country{China}
}
\email{guojian@idea.edu.cn}

% \author{Saizhuo Wang$^1$, Jian Guo$^2$ \\
%   $^1$ The Hong Kong University of Science and Technology \\
%   $^2$ IDEA Research\\
% }

\renewcommand{\shortauthors}{Bokai Cao et al.}

% \acmArticleType{Review}

% \author{{Anonymous}}
\begin{abstract}

Quantitative investment (quant) is an emerging, technology-driven approach in asset management, increasingly shaped by advancements in artificial intelligence. Recent advances in deep learning and large language models (LLMs) for quant finance have improved predictive modeling and enabled agent-based automation, suggesting a potential paradigm shift in this field. In this survey, taking alpha strategy as a representative example, we explore how AI contributes to the quantitative investment pipeline. We first examine the early stage of quant research, centered on human-crafted features and traditional statistical models with an established alpha pipeline. We then discuss the rise of deep learning, which enabled scalable modeling across the entire pipeline from data processing to order execution. Building on this, we highlight the emerging role of LLMs in extending AI beyond prediction, empowering autonomous agents to process unstructured data, generate alphas, and support self-iterative workflows.

\end{abstract}



\maketitle


% \thispagestyle{plain}
% \pagestyle{plain}

% Figures and tables are modularized for better modifications

\newcommand{\figureQuantPipeline}{
\begin{figure*}
    \centering
    \includegraphics[width=\textwidth]{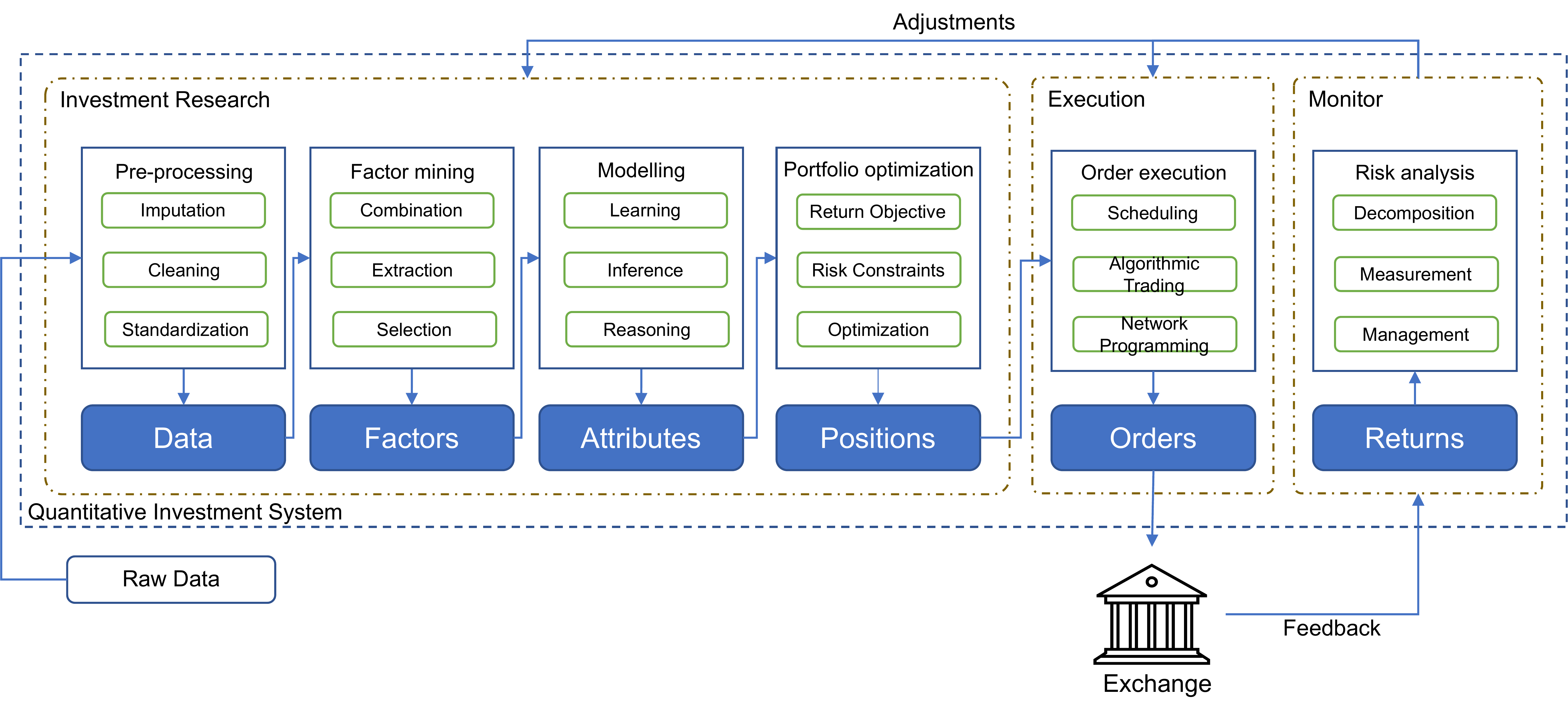}
    \caption{A typical pipeline of quantitative investment.}
    \label{fig:quant_pipeline}
\end{figure*}
}

\newcommand{\figureQuantWorkflow}{
    \begin{figure*}[!t]
        \centering
        \includegraphics[width=\textwidth]{mva_quant_workflow.pdf}
        \caption{A prototypical workflow of quantitative investment with comparisons between the current quantitative investment system (manual) and AI investment engineering (automatic).}
        \label{fig:quant_workflow}
    \end{figure*}
}

\newcommand{\figureMLQuantWorkflow}{
    \begin{figure}[!t]
        \centering
        \includegraphics[width=\textwidth]{quant_ml_workflow.pdf}
        \caption{ The quantitative investment workflow with machine learning. The upper part illustrates current practice, while the lower part illustrates the algorithmic design for AI investment engineering.}
        \label{fig:automl_quant_workflow}
    \end{figure}
}

\newcommand{\figureQuantSystem}{
    \begin{figure*}[!htbp]
        \centering
        \includegraphics[width=\textwidth]{quant_system.pdf}
        \caption{The architecture of a dedicated quantitative investment system.}
        \label{fig:quant_system}
    \end{figure*}
}

% Referenced figures for illustration
\newcommand{\figureNumericalData}{
    \begin{figure}
        \centering
        \begin{subfigure}{0.25\linewidth}
            \centering
            \includegraphics[width=\linewidth]{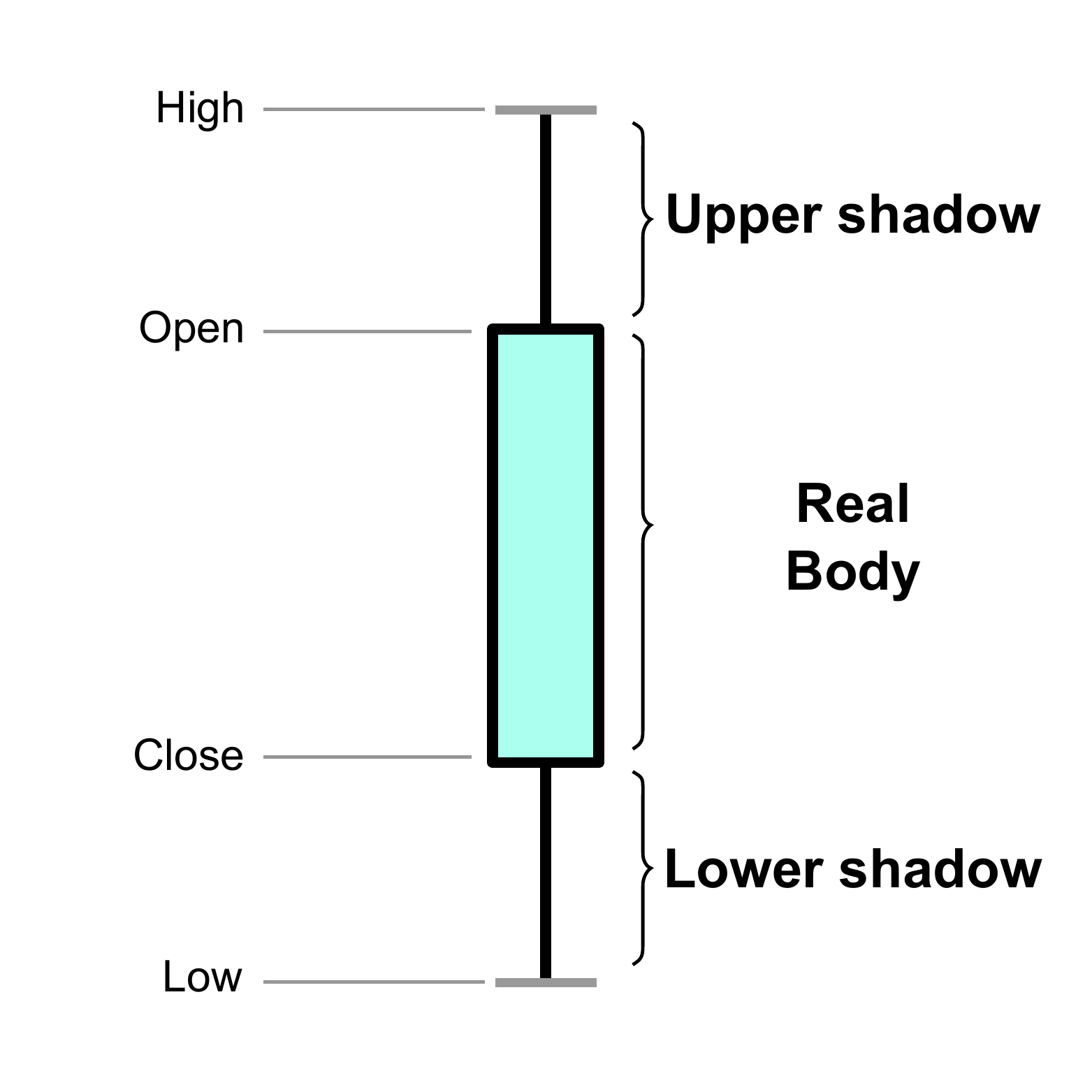}
            \caption{Candlestick chart\cite{noauthor_candlestick_2022} }
        \end{subfigure}
        \begin{subfigure}{0.3\linewidth}
            \centering
            \includegraphics[width=\linewidth]{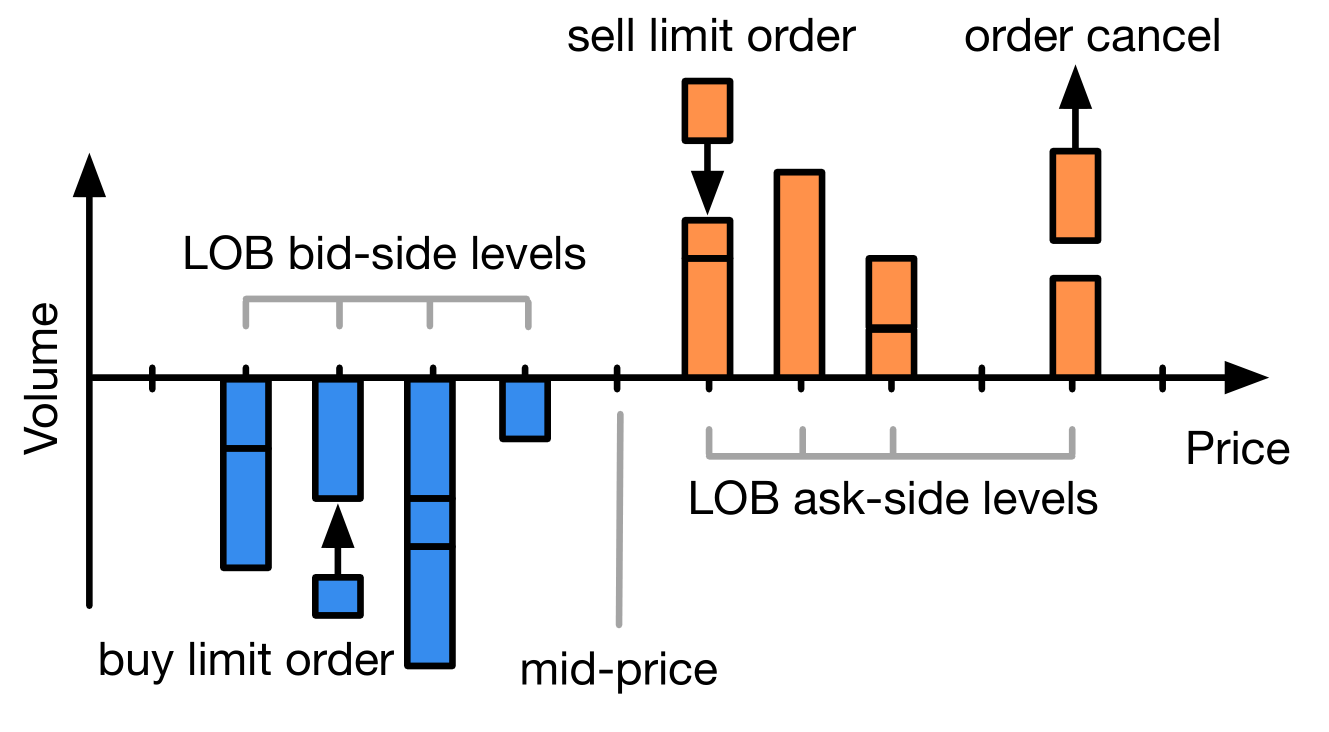}
            \caption{Limit Order Book\cite{wu_how_2021}}
        \end{subfigure}
        \begin{subfigure}{0.4\textwidth}
            \centering
            \includegraphics[width=\linewidth]{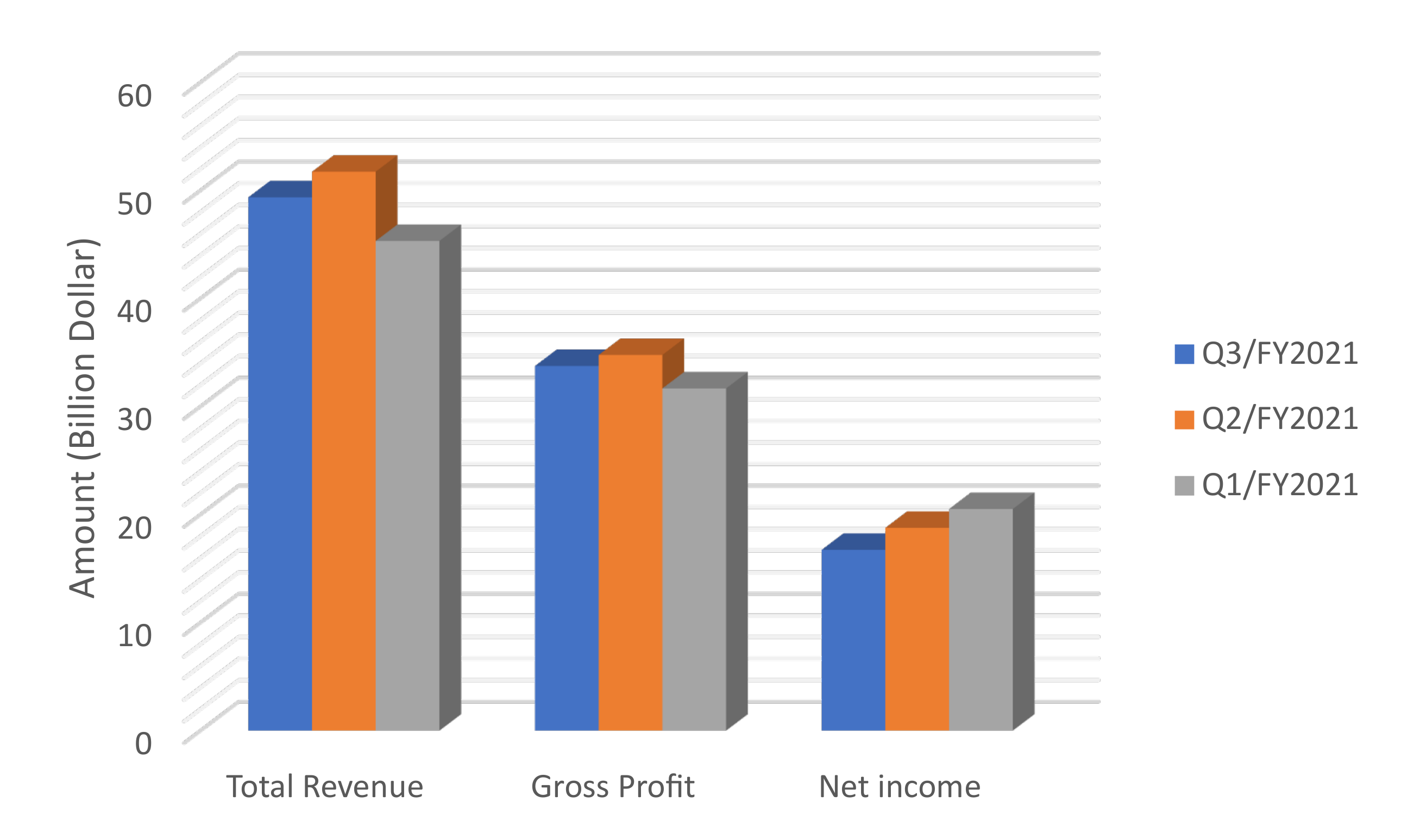}
            \caption{Fundamental data of Microsoft.}
        \end{subfigure}
        \caption{Examples of numerical data.}
        \label{fig:numerical_data}
    \end{figure}
}

\newcommand{\figureRelationalData}{
    \begin{figure}
        \centering
        \begin{subfigure}{0.46\textwidth}
            \centering
            \includegraphics[width=\linewidth]{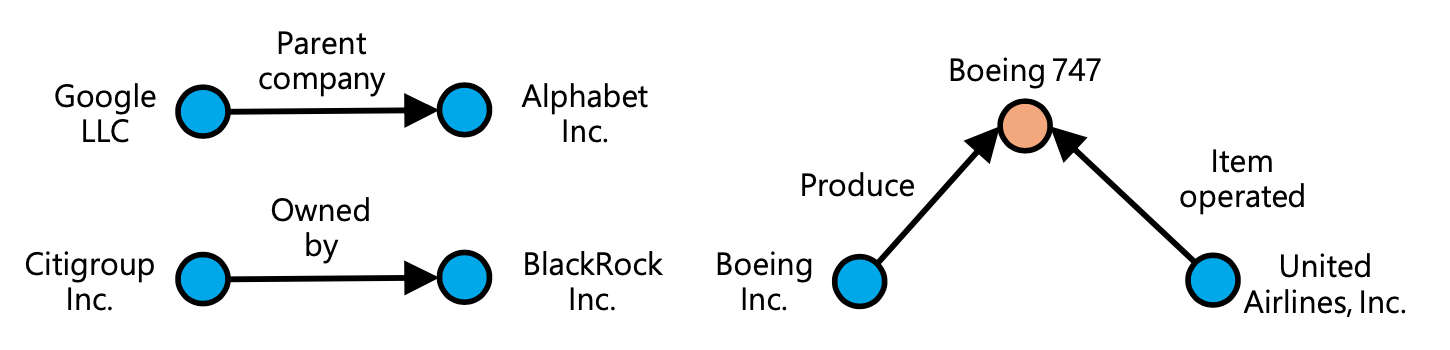}
            \caption{Pairwise edges \cite{feng_temporal_2019}}
        \end{subfigure}
        \begin{subfigure}{0.46\textwidth}
            \centering
            \includegraphics[width=\linewidth]{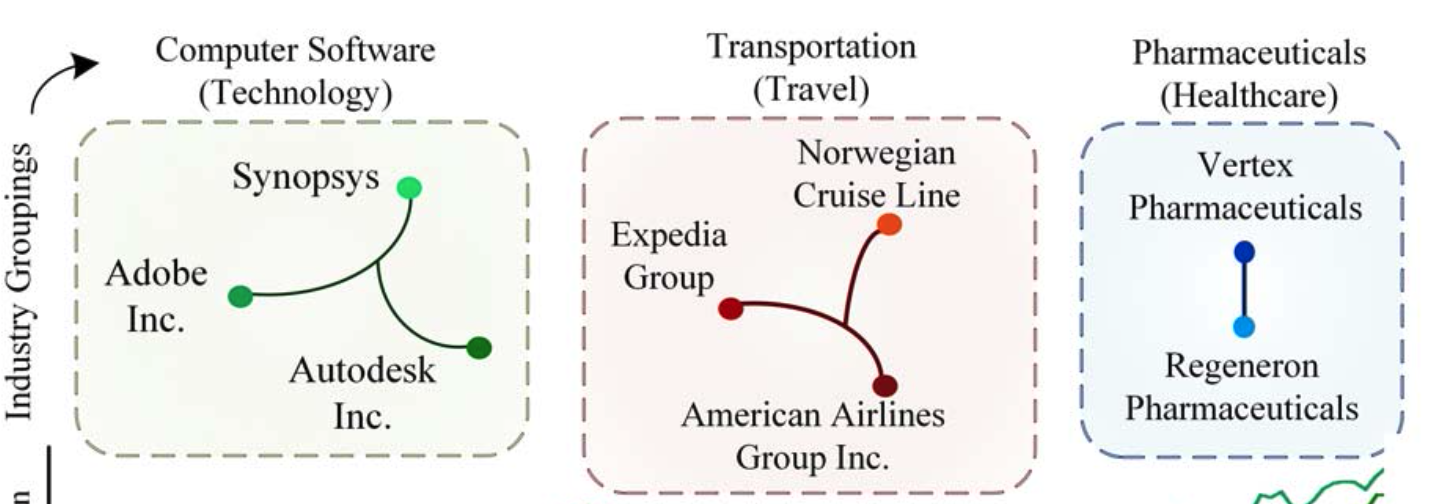}
            \caption{Hyperedges \cite{sawhney_spatiotemporal_2020}}
        \end{subfigure}
        \caption{Relational data examples.}
        \label{fig:relational_data}
    \end{figure}
}

\newcommand{\figureSpatiotemporal}{
    \begin{figure}
        \centering
        \begin{subfigure}{0.46\linewidth}
            \centering
            \includegraphics[width=\linewidth]{decoupled_spatiotemporal.png}
            \caption{Decoupled spatiotemporal model \cite{sawhney_spatiotemporal_2020}}
            \label{fig:spatiotemporal_decoupled}
        \end{subfigure}
        \begin{subfigure}{0.5\linewidth}
            \centering
            \includegraphics[width=\linewidth]{temporal_unrolling.png}
            \caption{An example of coupled spatiotemporal model using temporal unrolling \cite{kazemi_dynamic_2022}}
            \label{fig:spatiotemporal_coupled}
        \end{subfigure}
        \caption{Examples of spatiotemporal models.}
        \label{fig:spatiotemporal}
    \end{figure}
}

\newcommand{\figureRollingWindow}{
    \begin{figure*}[!htbp]
        \centering
        \includegraphics[width=\textwidth]{rolling_window.png}
        \caption{Illustration of the rolling window experiments \cite{kim_hats_2019}.}
        \label{fig:rolling_window}
    \end{figure*}
}

\newcommand{\figureEfficientFrontier}{
    \begin{figure*}[!htbp]
        \centering
        \includegraphics[width=0.46\linewidth]{Efficient Frontier.jpg}
        \caption{Efficient Frontier
        \cite{almgren_optimal_2000}.}
        \label{fig:efficient_frontier}
    \end{figure*}
}

\newcommand{\figureOverview}{
    \begin{figure*}[!htbp]
        \centering
        \includegraphics[width=1.1\linewidth]{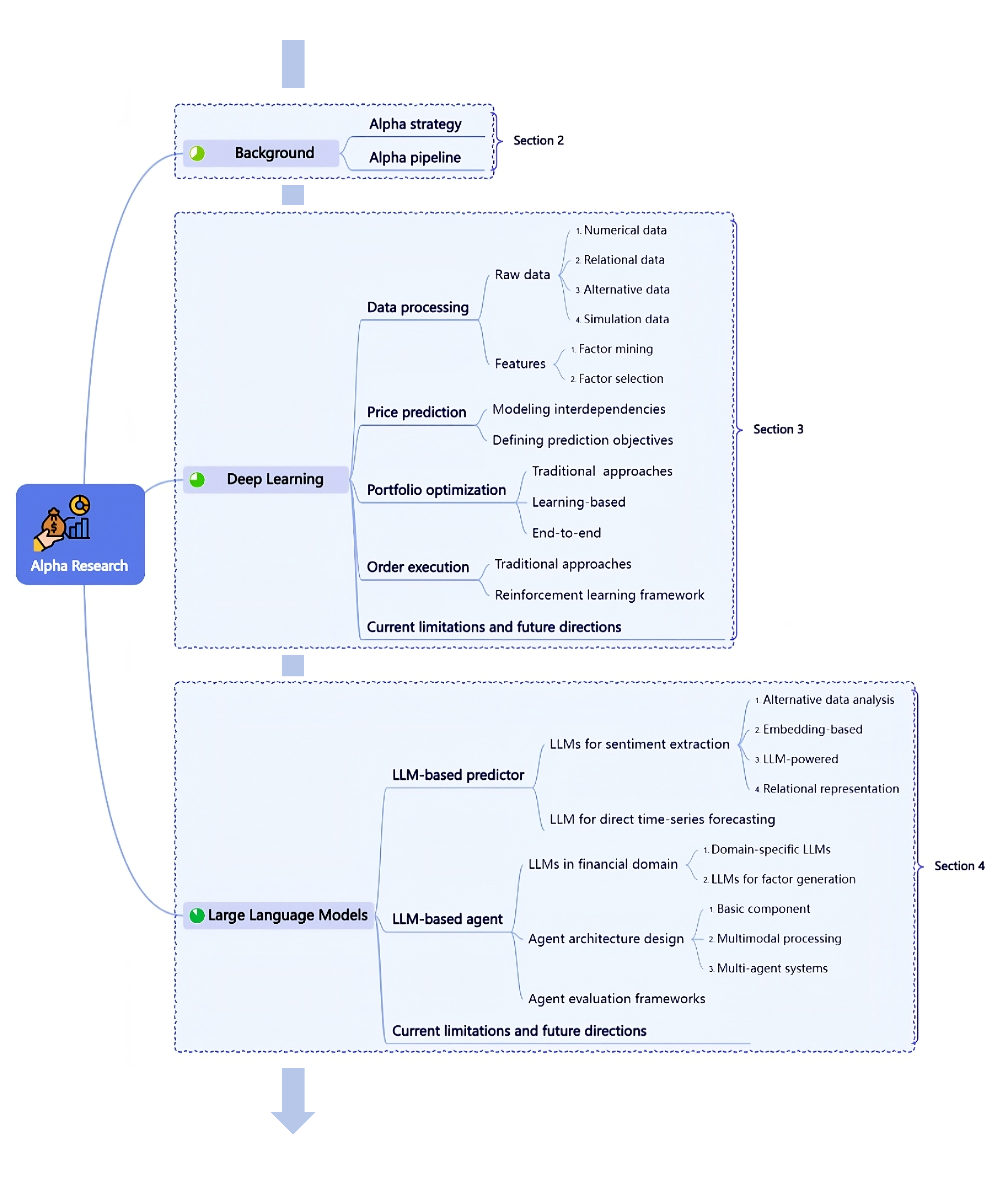}
        \caption{The overall framework of this paper.}
        \label{fig:overview}
    \end{figure*}
}

\newcommand{\figureStage}{
    \begin{figure*}[!htbp]
        \centering
        \includegraphics[width=0.8\linewidth]{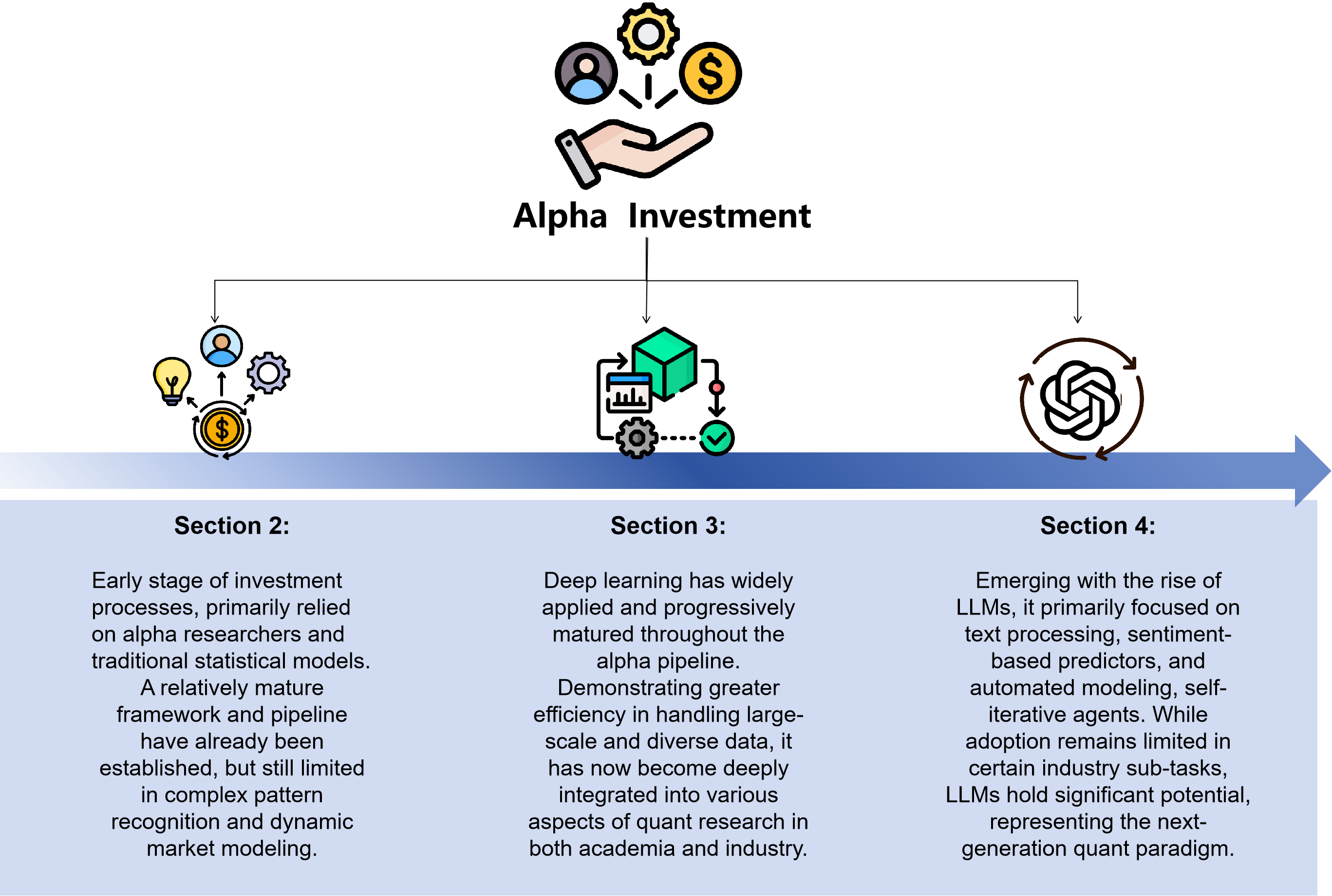}
        \caption{The evolutionary process of Alpha investment across different stages.}
        \label{fig:stage}
    \end{figure*}
}

\newcommand{\figureTemporalBlock}{
    \begin{figure*}[!htbp]
        \centering
        \includegraphics[width=0.8\linewidth]{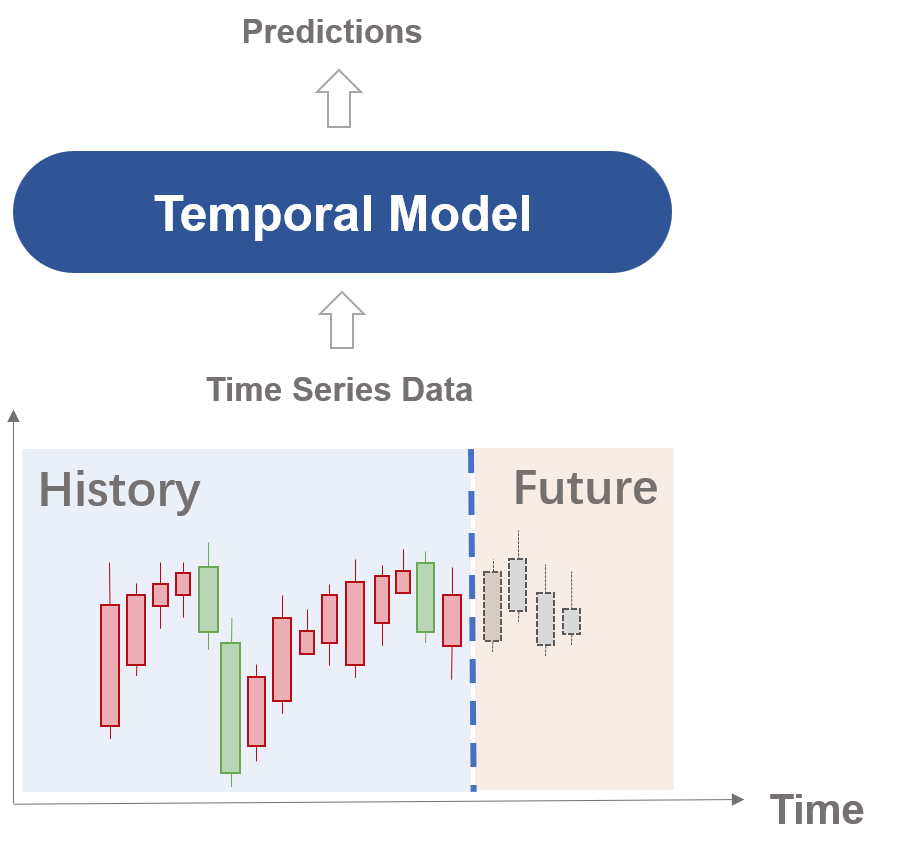}
        \caption{Temporal Patterns.}
        \label{fig:temporal_block}
    \end{figure*}
}

% Referenced figures for illustration
\newcommand{\figureDataDependencies}{
    \begin{figure}
        \centering
        \begin{subfigure}{0.3\linewidth}
            \centering
            \includegraphics[width=0.9\linewidth]{temporal_block.png}
            \caption{Temporal Patterns}
            \label{fig:temporal_patterns}
        \end{subfigure}
        \begin{subfigure}{0.3\linewidth}
            \centering
            \includegraphics[width=0.8\linewidth]{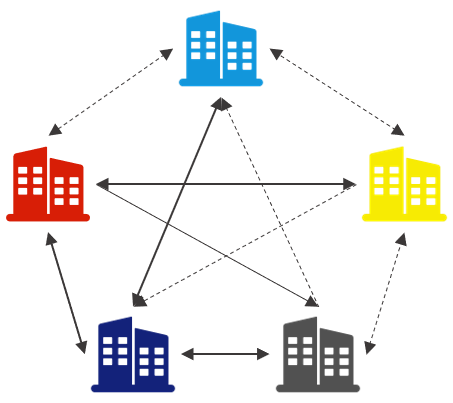}
            \caption{Spatial Patterns}
            \label{fig:spatial_patterns}
        \end{subfigure}
        \begin{subfigure}{0.3\textwidth}
            \centering
            \includegraphics[width=0.95\linewidth]{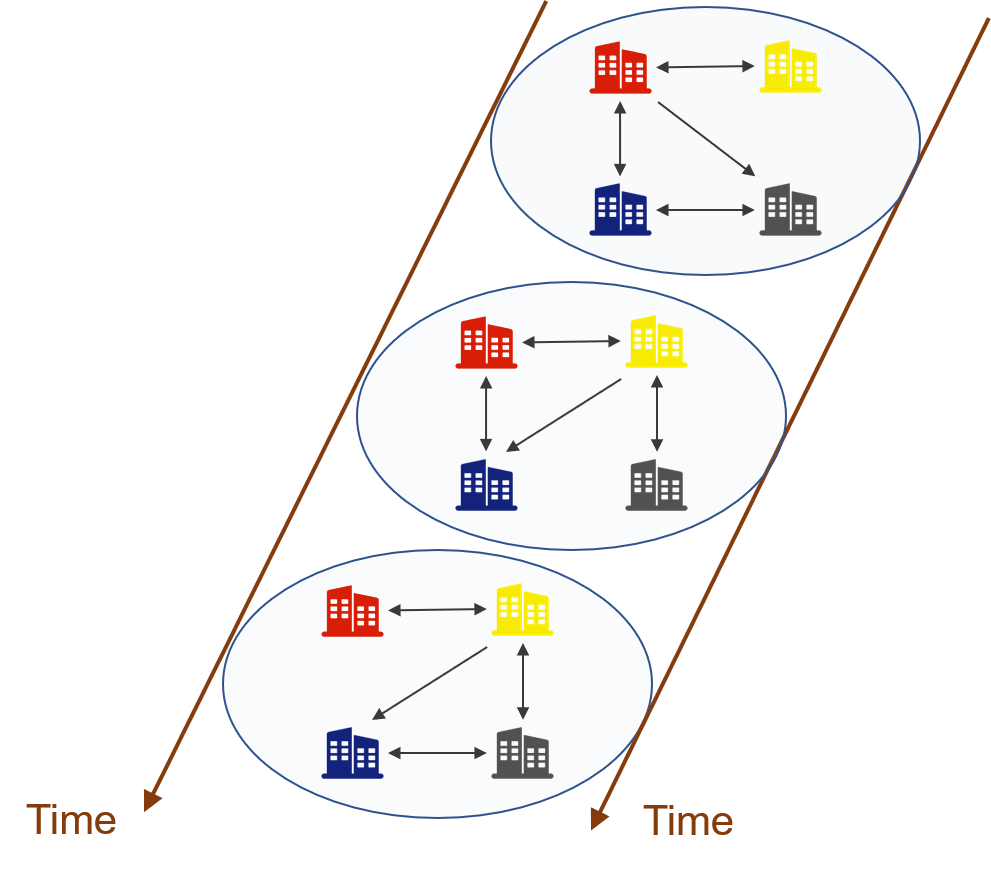}
            \caption{SpatioTemporal Patterns.}
            \label{fig:spatiotemporal_patterns}
        \end{subfigure}
        \caption{Modeling Data Interdependencies.}
        \label{fig:model_data_dependencies}
    \end{figure}
}

\newcommand{\tableSurveyComparison}{
\begin{table}[ht]
    \centering
    \begin{tabular}{cccc}
        \toprule
        \textbf{Work} & \textbf{Review} & \textbf{Perspective} & \\
         & 
    \end{tabular}
    \caption{Caption}
    \label{tab:my_label}
\end{table}
}

\newcommand{\tableQuantWorkflow}{
\begin{table*}[ht]
  \centering
  \caption{The input, output, goal, and typical methods of each stage in the quantitative investment workflow. \label{comparison_stages}}
  \resizebox{\textwidth}{!}{
    \begin{tabular}{|P{0.2\textwidth}|P{0.15\textwidth}|P{0.2\textwidth}|L{0.3\textwidth}|L{0.25\textwidth}|P{0.2\textwidth}|}
    \toprule
    \centering
    \textbf{Stage} & \textbf{Inputs} & \textbf{Outputs} & \multicolumn{1}{c|}{\textbf{Goals}} & \multicolumn{1}{c|}{\textbf{Typical Methods}} & \textbf{References} \\
    \midrule
    Pre-processing & Raw data & Data & Clean and transform the original data. & Batch processing & \cite{noauthor_pandas_2022} \\
    \midrule
    Factor mining & Data & Signal factors & Extract predictive factors.  & Human labor, heuristic search, machine learning & \cite{kakushadze_101_2016, cui_alphaevolve_2021, fang_alpha_2020} \\
    \midrule
    Asset selection & Signal factors & Predicted asset attributes & Make use of the factors to predict asset attributes such as return and volatility. & Machine learning, statistical modelling & \cite{zhang_stock_2017, lin_learning_2021} \\
    \midrule
    Portfolio optimization & Predicted asset attributes & Position assignments & Find the optimal asset allocation given the capital budget and return/risk expectations. & Numerical optimization, reinforcement learning & \cite{ye_reinforcement-learning_2020, xu_relation-aware_2020} \\
    \midrule
    Order execution & Position assignments & Orders & Convert the asset allocation weights to market orders. & Reinforcement learning & \cite{nevmyvaka_reinforcement_2006} \\
    \midrule
    Risk analysis & Market returns & Strategy adjustments & Measure the loss and volatility of the strategy based on the feedback and make adjustments to the strategy. & Machine learning, probabilistic modelling, statistical analysis & \cite{lin_deep_2021} \\
    \bottomrule
    \end{tabular}%
  }
  \label{tab:workflow_goals}%
\end{table*}%
}

\newcommand{\tableQuantTasks}{
\begin{table*}[ht]
  \centering
  \caption{Three prototypical tasks of quantitative investment.}
  \resizebox{\textwidth}{!}{
    \begin{tabular}{|P{0.2\textwidth}|L{0.4\textwidth}|L{0.4\textwidth}|P{0.2\textwidth}|}
    \toprule
    \centering
    \textbf{Tasks} & \multicolumn{1}{c|}{\textbf{Goals}} & \multicolumn{1}{c|}{\textbf{Profit}} & {\textbf{References}} \\ \midrule
    Trend trading & Predict whether the price of an asset/portfolio is going up or down. & The difference between buy and sell prices. & \cite{wang_alphastock_2019} \\
    \midrule
    Statistical arbitrage & Find correlated assets where there are price gaps. & Profit when the price gap is closed. & \cite{guijarro-ordonez_deep_2021, gatev_pairs_2006, rad_profitability_2016} \\
    \midrule
    Market making & Capture the micro price fluctuations in the market and provide enough liquidity. & The price difference between the sell and buy orders. & \cite{abernethy_adaptive_2013, zhong_data-driven_2020, gasperov_market_2021, haider_predictive_2022, spooner_robust_2020} \\
    \bottomrule
    \end{tabular}%
  }
  \label{tab:actual_tasks}%
\end{table*}%
}

\newcommand{\tableFundamentalData}{
\begin{table}[htbp]
  \centering
  \caption{An example of Microsoft's Fundamental data}
    \begin{tabular}{llll}
    \toprule
          & \textit{(Q3)FY2021} & \textit{(Q2)FY2021} & \textit{(Q1)FY2021} \\
    \midrule
    \textbf{Total Revenue} & 49.36B & 51.73B & 45.32B \\
    \textbf{Cost of revenue} & 15.62B & 16.96B & 13.65B \\
    \textbf{Gross profit} & 33.75B & 34.77B & 31.67B \\
    \textbf{Operating profit} & 20.36B & 22.25B & 20.24B \\
    \textbf{Net income} & 16.73B & 18.77B & 20.51B \\
    \bottomrule
    \end{tabular}%
  \label{tab:fundamental}%
\end{table}%
}

\newcommand{\tableDataCollection}{
\begin{table*}[!t]
  \centering
  \caption{Categorization of financial data. \qi{Removed the usage column. Reformat this table.} \qi{Add a comma for the sentences. Also add commas for other tables.}}
    \resizebox{\textwidth}{!}{
    \begin{tabular}{|L{0.15\textwidth}|L{0.15\textwidth}|L{0.2\textwidth}|L{0.4\textwidth}|L{0.3\textwidth}|L{0.15\textwidth}|}
    % \begin{tabular}{|P{0.1\textwidth}|P{0.1\textwidth}|P{0.08\textwidth}|P{0.2\textwidth}|P{0.1\textwidth}|P{0.2\textwidth}|P{0.1\textwidth}|}
    \toprule
    \multicolumn{3}{|c|}{\textbf{Modality}} & \multicolumn{1}{c|}{\textbf{Features}} & \multicolumn{1}{c|}{\textbf{Example}}  &  \multicolumn{1}{c|}{\textbf{References}} \\
    \midrule
    \multirow{3}{*}{Numerical} & \multirow{2}{*}{Quote data} & Regular interval & Quotes that are generated at regular time intervals & 1-minute candlestick chart &  \cite{sezer_financial_2020, wiese_quant_2020} \\ 
    \cmidrule{3-6}
    &  & Irregular interval & Quotes that are generated at irregular time intervals & Tick-level\footnote{\url{https://bit.ly/3Ob8Gjc}} order book data &  \cite{briola_deep_2020} \\
    \cmidrule{2-6}          
    & \multicolumn{2}{l|}{Fundamental data} & Fundamental data such as revenues and profits. & Financial statements &  \cite{tadoori_introduction_2020} \\
    \midrule
    \multirow{2}{*}{\parbox{0.08\textwidth}{Relational}} & \multicolumn{2}{l|}{Pairwise edges} & The relation between a pair of entities & Knowledge graph &   \cite{feng_temporal_2019, kim_hats_2019, xu_hist_2022, xu_rest_2021} \\
    \cmidrule{2-6}    
    & \multicolumn{2}{l|}{Hyperedges} & The relation involving a set of entities & Sector categorization &   \cite{sawhney_spatiotemporal_2020} \\
    \midrule
    \multirow{3}{*}{Alternative} & \multicolumn{2}{l|}{Text} & Information expressed in natural language. & Social media posts &  \cite{hu_listening_2019, li_modeling_2020, xu_stock_2018, sawhney_fast_2021} \\
    \cmidrule{2-6}
    & \multicolumn{2}{l|}{Images} & Images that are related to the traded asset & Satellite images & \cite{lutz_how_2018, partnoy_stock_2019} \\
    \cmidrule{2-6}          
    & \multicolumn{2}{l|}{Other modalities} & Anything & WiFi traffics, cell phone signals & \cite{jha_implementing_2018} \\

        \midrule
    \multirow{2}{*}{\parbox{0.08\textwidth}{Simulation}} & \multicolumn{2}{l|}{Time series} & Synthetic quote data & Simulated orders &   \cite{zhang2019deeplob, fu2019time, zhang2022data, li2024mars} \\
    \cmidrule{2-6}    
    & \multicolumn{2}{l|}{Tabular} & Database structured in a tabular form & Simulated financial statements &   \cite{fonseca2023tabular, surendra2017review, sattarov2023findiff} \\
        \cmidrule{2-6}    
        
    % & \multicolumn{2}{l|}{Unstructured data} & The relation involving a set of entities & Sector categorization &   \cite{sawhney_spatiotemporal_2020} \\
    
    \bottomrule
    \end{tabular}%
    }
  \label{tab:data_collection}%
\end{table*}%
}

\newcommand{\tableModelArch}{
\begin{table*}[!htbp]
 \centering
 \caption{Summary and comparison of different modelling block choice for the search space of model generation}
 \resizebox*{\textwidth}{!}{
   \begin{tabular}{|L{0.1\textwidth}|L{0.08\textwidth}|L{0.1\textwidth}|L{0.12\textwidth}|L{0.2\textwidth}|L{0.2\textwidth}|L{0.08\textwidth}|}
   \toprule
   \multicolumn{2}{|L{0.15\textwidth}|}{\textbf{Type}} & \textbf{Work with} & \textbf{Capability} & \textbf{Pros}  & \textbf{Cons}  & \textbf{References} \\
   \midrule
   \multirow{3}{*}{Temporal} & Regular Time Series Model   & \multirow{2}{*}{\parbox{0.1\textwidth}{Time Series Data}} & \multirow{2}{*}{\parbox{0.12\textwidth}{Easy modelling for sequences where all timesteps have same weights}} & Can look back on the whole history & Forgetting, Gradient Issues, Slow run time & \cite{qin_dual-stage_2017} \\
   \cmidrule{2-2}\cmidrule{5-7}          
   & Irregular Time Series Model &       &       & Flexible application scenarios that do not require equal time intervals & Higher computation cost & \cite{rangapuram_deep_2018} \\
   \cmidrule{2-2}\cmidrule{5-7}          
    & 1D-CNN &       &       & Efficiently aggregate information for long steps & Weaker representation capability &  \\
   \midrule
   \multirow{2}{*}{Spatial} & Dense Spatial Model & \multirow{2}{*}{\parbox{0.1\textwidth}{Cross-Sectional Data}} & \multirow{2}{*}{\parbox{0.12\textwidth}{Capture relations among entities at some fixed time step}} & Easy to model, Directly captures spatial relations & Limited receptive fields, Hard to train because of high noise ratio & \cite{wang_alphastock_2019} \\
   \cmidrule{2-2}\cmidrule{5-7}          
   & Sparse Spatial Model   &       &       & The relations act as a filter for the useful information so that the spatial information is `purified' and the training efficiency is thus increased & The graph structure itself requires more work to produce. GNN faces issues like over-smoothing & \cite{feng_temporal_2019, kim_hats_2019} \\
   \midrule
   \multirow{2}{*}{Spatiotemporal} & Decoupled &   \multirow{2}{*}{\parbox{0.1\textwidth}{Multiple time-series data with interrelations}} & \multirow{2}{*}{\parbox{0.12\textwidth}{Captures information both on the time axis and the space axis.}} & Easy to model. Components are not tightly coupled & The two axes are still disconnected. May miss some cross-axis information. &  \cite{liu2019hyperbolic}\\
   \cmidrule{2-2}\cmidrule{5-7}          
   & Coupled &       &       & Fuses information on both side. & Modelling Difficulty. May require more computing power. & \cite{chen_tamp-s2gcnets_2021} \\
   \bottomrule
   \end{tabular}%  
 }
 \label{tab:models}%
\end{table*}%
}

\newcommand{\tableEvalMetrics}{
\begin{table*}[!t]
  \centering
  \caption{A summary of evaluation metrics. \qi{reformat this table}}
  \resizebox*{\textwidth}{!}{
    \begin{tabular}{|L{0.2\textwidth}|L{0.2\textwidth}|L{0.35\textwidth}|L{0.35\textwidth}|L{0.15\textwidth}|}
      \toprule
      \multicolumn{2}{|c|}{\textbf{Metrics}} & \multicolumn{1}{c|}{\textbf{Examples}} & \multicolumn{1}{c|}{\textbf{Motivations}}  & \multicolumn{1}{c|}{\textbf{References}} \\
      \midrule
      \multirow{2}{*}{\parbox{0.15\textwidth}{Machine learning metrics}} & Prediction-based metrics & Root mean squared error (RMSE), area under the ROC curve (AUC), F1 Score & Evaluating the model with prediction metrics. & \cite{li_modeling_2020, wang_clvsa_2019} \\ 
      \cmidrule{2-5}          
      & Information retrieval-based metrics & Precision and recall, mean reciprocal rank (MRR), normalized discounted cumulative gain (NDCG) & Evaluating the model with retrieval metrics. & \cite{feng_temporal_2019, wang_alphastock_2019} \\
      \midrule
      \multirow{2}{*}{\parbox{0.15\textwidth}{Investment-specific metrics}} & Correlation metrics & Information coefficient, information ratio & Calculating the correlation with the ground truth. & \cite{xu_hist_2022} \\
    \cmidrule{2-5}
    & Portfolio metrics & Annualized return, maximum drawdown, Sharpe ratio & Evaluating model profitability. & \cite{kim_hats_2019, wang_alphastock_2019} \\
      \bottomrule
      \end{tabular}%  
  }
  \label{tab:eval_metrics}%
\end{table*}%
}

\newcommand{\tableTrainingMethods}{
\begin{table*}[htbp]
 \centering
 \caption{Summary of training methods}
 \resizebox*{\textwidth}{!}{
   \begin{tabular}{|L{0.1\textwidth}|L{0.18\textwidth}|L{0.08\textwidth}|L{0.18\textwidth}|L{0.18\textwidth}|L{0.1\textwidth}|}
     \toprule
     \textbf{Methods} & \textbf{Scenarios} & \textbf{Example} & \textbf{Pros}  & \textbf{Cons}  & \textbf{References} \\
     \midrule
     Supervised Learning & The goal of the task can be represented by some strong supervision signal. & Stock Prediction & The direct supervision can provide clear guide to train the model. & The learning goal is usually inconsistent with the final goal. & \cite{hu_survey_2021, yoo_accurate_2021, jiang_applications_2021, feng_enhancing_2019, ding_hierarchical_2020, ding_knowledge-driven_2016, liu_transformer-based_2019, chen_trimming_2020} \\
     \midrule
     Unsupervised Learning & We want to capture the intrinsic features of the data and this is usually done in a generative way & Financial Time Series Synthesis & No need for labelling and annotations. Can perform generations. & Weak training convergence guarantee & \cite{wiese_quant_2020} \\
     \midrule
     Reinforcement Learning & It is hard to find a direct supervision (i.e. ground truth) for the task and the data itself do not contain much meaningful patterns for self-supervision & Order Execution & Ease of modelling: Almost all the problems can be modelled with RL.\newline{}No need to find ground truth labels, thus preventing the prediction-reality gap & Hard to converge: It is widely acknowledged that RL models are unstable and hard to train. & \cite{spooner_robust_2020, yang_deep_2020, zhong_data-driven_2020} \\
     \bottomrule
   \end{tabular}%  
 }
 \label{tab:training_methods}%
\end{table*}%
}

\newcommand{\tableLibraryComparison}{
\begin{table*}[!htbp]
  \centering
  \caption{Library Summary}
    \begin{tabular}{|c|p{8.835em}|c|c|c|c|}
    \toprule
    \multicolumn{2}{|p{15.25em}|}{Library} & \multicolumn{1}{p{7.665em}|}{vn.py} & \multicolumn{1}{p{5em}|}{qlib} & \multicolumn{1}{p{5em}|}{QuantLib} & \multicolumn{1}{p{9.5em}|}{FinRL} \\
    \midrule
    \multicolumn{2}{|p{15.25em}|}{Developer} & \multicolumn{1}{p{7.665em}|}{VeighNa} & \multicolumn{1}{p{5em}|}{Microsoft} & \multicolumn{1}{p{5em}|}{QuantLib Community} & \multicolumn{1}{p{9.5em}|}{AI4Finance Foundation} \\
    \midrule
    \multicolumn{2}{|p{15.25em}|}{License} & \multicolumn{1}{p{7.665em}|}{MIT} & \multicolumn{1}{p{5em}|}{CLA} & \multicolumn{1}{p{5em}|}{BSD} & \multicolumn{1}{p{9.5em}|}{MIT} \\
    \midrule
    \multicolumn{1}{|c|}{\multirow{4}[8]{*}{Data}} & Data Source &       & \multicolumn{1}{p{5em}|}{Yahoo! Finance} &       & \multicolumn{1}{p{9.5em}|}{Yahoo! Finance, WRDS} \\
\cmidrule{2-6}          & Data Type &       & \multicolumn{1}{p{5em}|}{Stocks} & \multicolumn{1}{p{5em}|}{Stocks, Forex, Futures, Options} &  \\
\cmidrule{2-6}          & Supported Data Modality &       & \multicolumn{1}{p{5em}|}{Quote Data, Relational Data} &       &  \\
\cmidrule{2-6}          & Factor Engine &       & \multicolumn{1}{p{5em}|}{abc} &       &  \\
    \midrule
    \multicolumn{1}{|c|}{\multirow{3}[6]{*}{Modelling}} & Supported Algorithm Types &       &       &       &  \\
\cmidrule{2-6}          & Backend Engine &       &       &       &  \\
\cmidrule{2-6}          & GPU Acceleration &       &       &       &  \\
    \midrule
    \multicolumn{1}{|c|}{\multirow{3}[6]{*}{Evaluation}} & Backtest Engine &       &       &       &  \\
\cmidrule{2-6}          & Analysis Engine &       &       &       &  \\
\cmidrule{2-6}          & Visualization &       &       &       &  \\
    \midrule
    \multicolumn{1}{|c|}{\multirow{3}[6]{*}{Deployment}} & Database Engine Support &       &       &       &  \\
\cmidrule{2-6}          & Trading API &       &       &       &  \\
\cmidrule{2-6}          & Exchange Interface &       &       &       &  \\
    \bottomrule
    \end{tabular}%
  \label{tab:lib_summary}%
\end{table*}%
}

% Main Article
\section{Introduction \label{sec:intro}}
% Quantitative investment strategies, leveraging statistical analysis, optimization techniques, with increasingly, deep learning (DL) and large language model (LLM) algorithms, represent a sophisticated facet of the financial industry. These approaches utilize mathematical models and computational powers to identify market inefficiencies and guide investment decisions, benefiting from the exponential growth in data availability and technological advancements. Stock alpha strategies, which aim to generate returns exceeding the market benchmark by exploiting mispricings, exemplify the successful application of quantitative methods. The integration of quantitative techniques in these strategies has shown promise in unveiling complex patterns within financial data, offering new pathways for alpha generation. This paper examines the application of DL and LLM in quantitative investment, focusing on stock alpha strategies, and reviews the existing literature while highlighting the challenges and potential paradigm shift through LLM-based approaches.

% First point: Asset mgmt, quant, and stock alpha

Asset management is a crucial and expanding segment of the financial industry, with \textbf{Quantitative Investment (Quant)} emerging as a key approach within it. Quantitative investment strategies leverage statistical analysis, optimization techniques, and increasingly, AI algorithms to identify and exploit market inefficiencies. Benefiting from the exponential growth in data availability, computational power, and technological innovations, these approaches significantly improve investment decision-making and provide a competitive edge in the financial market.

Among various quantitative investment approaches, \textbf{alpha strategy} has received considerable attention for its strong capacity to capture market inefficiencies and its natural alignment with AI-driven predictive methods. The pursuit of 'alpha' refers to predicting individual asset's excess returns over the market's overall performance, such as a stock index, and is the central focus of portfolio managers. The development of alpha strategies typically includes four steps: data processing, model prediction, portfolio optimization, and order execution (as introduced in subsection \ref{alpha_pipelne}). These four sub-tasks, though distinct, are closely interconnected, all working towards the common goal of maximizing excess returns while controlling risks. Compared to other quantitative investment strategies, such as high-frequency trading or arbitrage, alpha strategies have been shown to have great capacity and effectiveness by exploiting market mispricings. As a result, alpha strategies receive the highest attention, research focus, and market share of researchers and investors, representing the core technology in quantitative investment. In this survey, we take the alpha strategy as a representative example of quantitative investment and center our discussion on how AI plays a role in this field.

In recent years, the application of \textbf{deep learning (DL)} techniques in alpha strategies has shown promising results, demonstrating the ability to identify complex patterns and relationships in financial data that are difficult to detect using traditional quantitative methods. Meanwhile, \textbf{large language models (LLMs)}, such as GPT-series \cite{achiam2023gpt}, BERT \cite{devlin2018bert}, and their financial variants, have shown remarkable power in understanding contextual data, generating accurate interpretations, and reasoning like human analysts. Therefore, their application in finance, particularly in quantitative investment, has sparked endless possibilities.

This paper focuses on the evolution, application, and respective advantages of DL and LLMs in quantitative investment, with a specific emphasis on alpha strategies, providing a comprehensive review of the existing literature, and discussing the potential, challenges, and limitations of how LLMs could enhance DL-based approaches.

\subsection{Evolution of Alpha Strategy Investment}
% Second point: from manual modelling to deep learning to LLM

\figureStage
The evolution of the alpha strategy could been characterized by a three-stage progression from manual labeling of trading signals to the use of deep learning models and ultimately to an era of agent interaction and decision-making between LLM agents (Figure \ref{fig:stage}). In the early stages, the focus was on traditional statistical modeling of market patterns, relying on the expertise of individual researchers to identify profitable trading signals and develop corresponding models. However, this approach had limitations due to the complexity of financial markets and the difficulty of capturing all relevant factors in a model. It still relies heavily on the skill and experience of human researchers to evaluate and execute trading strategies.

As the field has matured, the application of deep learning has opened up new possibilities for quantitative investment. Particularly in the context of alpha research, deep learning has been shown to be effective in identifying inherent patterns. For example, deep learning models have been used to analyze factors such as spatial interconnectedness \cite{wang2021review}, long-term temporal dependence \cite{zhang2022transformer}, and news sentiment \cite{jin2020stock} to predict the price movements and manage positions. While the use of deep learning holds great promise, there are also challenges that need to be addressed. One major challenge is the risk of overfitting, which can lead to poor performance when the model is applied to new data. Another challenge lies in improving interpretability and accuracy to further understand, reason, and interact with vast volumes of multimodal data. Despite these challenges, the application of deep learning in quantitative investment is expected to continue to grow in both academia and industry, as investors seek to gain an edge in a competitive market.

More recently, LLMs have emerged as a powerful tool in quantitative investment, characterized by rapid development and enormous application potential, attracting significant attention. They excel in understanding and processing multimodal data and possess the potential to autonomously handle complex tasks of reception, comprehension, and inference over large-scale datasets. In current mainstream research, LLMs primarily serve two roles in alpha strategy: as predictors (subsection ~\ref{llm:pred}) and as agents (subsection ~\ref{llm:agent}), handling various tasks. In both cases, they contribute high-level insights built upon deep learning frameworks and have the potential to further evolve AI-powered deep learning investment methods into the AI-automated stage. However, the practical deployment of LLMs is still in its early stages. We will highlight their current limitations and discuss potential future directions for their development (subsection ~\ref{llm:limit}).
\figureOverview

% Third point: The motivation for this survey
\subsection{Motivation and Contribution of this Survey}

The use of deep learning and LLMs in alpha strategies of quantitative investment has recently seen a surge of interest, with many studies focusing on the application in various aspects of the alpha research pipeline. However, most of these studies are relatively isolated in specific tasks or disciplines, and there is a lack of a unified view of the whole landscape of quantitative investment, particularly in the context of alpha strategy. This survey will also systematically summarize the evolution of alpha research from the perspective of phased algorithmic evolution. Additionally, quant is a field where research and practice are highly interconnected, while existing survey papers are limited to filling technical gaps between practical opportunities and research theories related to the combination of LLMs and DL-based alpha models. There's a lack of a comprehensive framework and forward-looking perspective on the future of deep learning and LLMs research workflows from a real-world viewpoint.

To address these issues, this survey paper aims to provide readers with a more integrated and comprehensive view of the alpha strategy. We intend to achieve this by surveying relevant works that cover all types of deep learning and LLMs tasks in the whole alpha pipeline, offering a holistic and interconnected view of the field. Moreover, our survey paper seeks to provide a broader research perspective by starting from real-world applications, highlighting the practical issues and challenges faced by investors  to reveal the generic research problems and promote future studie. The overall framework of this paper is shown as Figure \ref{fig:overview}.

Key contributions of this paper include:

\begin{itemize}
    \item It provides a comprehensive survey of the existing research on the use of deep learning and large language models in alpha strategies, connecting different works in a concrete research pipeline. This survey is the first of its kind to cover the topic comprehensively, offering a holistic view of the field.
    \item It Introduces the domain from an interdisciplinary standpoint, emphasizing practical applications to derive key research questions for quantitative investigations. Additionally, it discusses the most challenging problems from a practical perspective and offers insights into potential future research directions.
    
    \item It systematically compares the technical approaches, strengths, and weaknesses across the three stages of quantitative investment: traditional statistical models, DL-based methods, and LLM-based approaches. Building on the iterative development, it identifies key gaps and leads alpha strategies to the next stage.
\end{itemize}

\section{Background}
In this section, we provide a brief overview of alpha strategy and its role in quantitative investment. Specifically, we begin by introducing alpha strategies and then discuss the alpha pipeline in quant, which serves as the framework for developing and implementing the investment process.

\subsection{Alpha Strategy}
Alpha strategies are investment approaches that focus on identifying opportunities yielding returns that exceed the market benchmark by exploiting inefficiencies or mispricings. These strategies typically consist of two components: the alpha side and the hedging side.

The alpha side of an alpha strategy focuses on generating excess returns by identifying profitable investment opportunities. To achieve this, it aims to make accurate predictions about the trends of individual instruments, sectors, or the overall market. These predictions are then used to allocate capital to different investment instruments to maximize returns. The alpha side faces several common challenges. For example, predicting asset prices involves handling high-dimensional, noisy data with complex patterns, while portfolio allocation requires optimizing the trade-offs between risk and return. Deep learning techniques have been shown to be effective in addressing these challenges, which is why recent works have applied deep learning to alpha strategies.

The hedging side of an alpha strategy focuses on managing the risks associated with market movements. Its goal is to mitigate market risks, ensuring that the excess returns generated by the alpha side are not eroded. To achieve this, the hedging side typically employs hedging portfolios, such as derivatives like stock index futures or options, while aiming to minimize hedging costs. The hedging side also faces several common challenges. For example, selecting the appropriate hedging portfolio requires understanding the relationship between the portfolio and the underlying investments. Additionally, hedging portfolios may incur transaction costs, which can affect returns.

While both the alpha side and the hedging side of an alpha strategy aim to generate returns while controlling risks, they face different challenges and require distinct solutions. In this paper, we primarily focus on the alpha side of alpha strategies and discuss recent works that have employed deep learning techniques and large language models to address the challenges faced by the alpha side. However, it is important to note that the hedging side is also an essential component of alpha strategies, and interested readers can refer to relevant literature for further study.

\figureQuantPipeline

\subsection{Alpha Pipeline}
\label{alpha_pipelne}
In its essence, an alpha strategy aims to identify which instruments to trade, and actually trade them to earn profits. In its most simple form, this idea can be practiced in a one-step way. The investor can just look at the market, find some assets he wants to trade, and put the corresponding orders on the market.

However, things become much more complex when the capacity of the strategy grows in terms of both the amount of money and the number of assets involved in the investment. In this case, trading decisions cannot be carried out just with simple human operations. Instead, the whole strategy needs to be standardized into several sub-tasks, where each task is well-formulated and has a well-developed toolbox to deal with it. As shown in Figure \ref{fig:quant_pipeline}, an alpha strategy can be decomposed into a pipeline consisting of several sub-tasks, which we will illustrate in the following.
\begin{itemize}[leftmargin=*]
    \item \textbf{Data processing}: The whole pipeline starts from analyzing the data from the market, and the data come in various forms. To perform modern data mining techniques on these data, the data must be first cleaned and standardized into unified forms. The pre-processing step is hence involved to do the cleaning, standardization, and imputation of data. Meanwhile, pre-processed data can developed into features to better integrate market information and serve as input for the next stage of the model.
    \item \textbf{Model prediction}: Although investments can be done with various motivations, the most important one should be the expected future performance of the asset price. Therefore, to form an investment decision, we need to first predict the future of the assets of interest, and then make decisions accordingly. Price prediction aims to make predictions about assets, such as their future price change, volatility, etc. This prediction task is what deep learning is good at, so a bunch of deep learning techniques have been studied in making better and more accurate predictions.
    \item \textbf{Portfolio optimization}: Model predictions themselves cannot be directly used as investment decisions. They are utilized in the portfolio optimization stage to generate investment decisions. Essentially, the portfolio optimization step takes the various kinds of market predictions (e.g. price change, volatility) as input, and outputs the corresponding investment decisions such as the amount of money allocated for each asset in the next holding period (also known as \textit{positions}). The core problem in this stage is to find an optimal portfolio optimization that maximizes the objective defined by investors such as maximizing the expected risk-adjusted returns, subject to the constraints defined simultaneously, such as the maximum volatility constraint or the diversity constraint defined for risk control. Traditionally, this problem is formulated as an optimization problem and the corresponding optimization techniques can be applied to solve this problem.
    \item \textbf{Order execution}: The portfolio allocations need to be implemented by actually putting orders on the market and making the deal. And this order execution step is by no means an easy task for alpha strategies that involve a large amount of money. This is because orders placed on the market will inevitably bring fluctuations on the market, driving the price to the opposite direction that is beneficial for the investor. And when the trade volume becomes large, this effect is also magnified and might introduce great loss to the strategy. Hence, the order execution stage is involved to minimize such loss by splitting big orders into smaller ones and deciding the appropriate time to execute them. Traditionally, this problem has been formulated as an optimal control problem where the dynamics of the limit order book are extensively studied.
    % \item \textbf{Risk management}: The investment pipeline is not an open-loop system. Instead, it needs feedback from the market to adjust its behaviors. The risk management stage is therefore involved in the whole pipeline to take feedback signals from the market and make corresponding adjustments to the previous signals. For example, the risk management module may calculate the risk exposure of the current portfolio to different sectors, and make balancing or neutralization adjustments by injecting constraints or adjusting the objective functions in the portfolio optimization module.
\end{itemize}

The investment pipeline is not an open-loop system. Instead, it needs feedback from the market to adjust its behaviors. The monitor module is therefore involved in the entire pipeline to conduct risk analysis, collect feedback signals, and make corresponding adjustments to the previous processes. For example, the risk management module calculates the risk exposure of the current portfolio to different sectors. Then, from the perspective of the overall process, it makes balancing or neutralization adjustments by injecting constraints or adjusting the objective functions in the portfolio optimization module.

\section{Deep Learning in Alpha pipeline}

Deep learning has been widely applied throughout the whole alpha pipeline. In this section, we systematically analyze how deep learning is utilized to enhance traditional alpha research at each sub-task of the previously discussed pipeline.

\subsection{Data Processing}

\tableDataCollection
\figureNumericalData
As stated before, the data used for alpha research need to be first pre-processed into unified format, and then used for prediction. Here we divide data before and after pre-processing into two types, namely raw data, and features.

\subsubsection{Raw Data}
Financial markets constantly generate heterogeneous data with various modalities. Based on their source and modality, we categorize financial data used in quant strategies as numerical data, relational data, alternative data and simulation data. A summary and a comparison of different types of data are presented in Table \ref{tab:data_collection} and we will elaborate on each of them in the following.

\paragraph{Numerical data}
Numerical data are the most widespread type of data in the financial world, it can be categorized as quote data that indicate the movement of asset prices of arbitrary frequency, and fundamental data that reflect the operations of the underlying economic entities of securities. 

Quote data typically includes candlestick chart and limit order book, as illustrated in Figure \ref{fig:numerical_data}. Candlestick chart is a way to illustrate the price movement of a financial instrument, which consists prices at four different information dimension in an time interval, namely the open, close, high, and low prices. A limit order book refers to an electronic list of buy and sell limit orders organized by price levels. Fundamental data is widely used by analysts to determine the intrinsic value of a financial instrument. Important sources of fundamental data are balance sheets, income statements, and cash flow statements from financial statements.

% Financial markets are intrinsically dynamic, and it is this constant change that is meaningful for analysis. Hence, numerical data in practice are often treated as time series, from which useful patterns are recognized and exploited. Under such circumstances, a very important property of numerical data is their frequency, i.e., the interval at which new data points are generated. Generally speaking, fundamental data have the lowest frequency, changing every few months. In contrast, quote data such as candlestick charts have frequencies at multiple levels, ranging from the lower ones such as week- or day-level frequencies to the higher ones such as minute- or second-level frequencies. More extremely, the quote data with the maximum temporal resolution, namely tick-level data, records the history of each order and transaction sequentially. Compared with other types of time series data with fixed interval, tick-level data are irregular time-series with non-uniform interval, requiring special data mining techniques on them.

Financial markets are intrinsically dynamic, numerical data are often treated as time series. Under such circumstances, a very important property is their frequency, i.e., the interval at which new data points are generated. Generally speaking, fundamental data have the lowest frequency, changing every few months. In contrast, quote data have frequencies at multiple levels, ranging from the lower ones such as week- or day-level to the higher ones such as minute- or second-level frequencies. More extremely, the quote data with the maximum temporal resolution, namely tick-level data, records the history of each order and transaction sequentially. 

\figureRelationalData

\paragraph{Relational data}
While numerical data describes individual financial entities, the relationships between them can also impact market trends. We refer to such data as \textit{relational data}, which describes the ubiquitous relationships between two or more financial entities. Formally, relational data are usually represented as a graph $\mathcal{G} = V \times E$, where $V$ is the set of nodes and $E$ is the set of edges. The edges can be either \textit{pairwise edges} or \textit{hyperedges} \cite{hyperedge_wolfram}, based on the number of entities involved in the relations. In practice, relational data have rich semantics, ranging from the business relationships between companies, to various events involving financial entities, to the correlations and causal relationships between entities from a statistical perspective. An illustration of pairwise edges and hyperedges is present in Figure \ref{fig:relational_data}. 
\begin{enumerate}[wide]
    \item Pairwise edges: As the most common type of relations, pairwise edges describe the relations between a pair of entities. A pairwise edge can be represented as a triple $(v_1, r, v_2)$ where $v_1, v_2$ are entities and $r$ is the edge type or edge weight. Meanwhile, it can be further categorized into static and dynamic depending on whether the underlying relationship changes over time. Static edges usually include the stable relationships such as upstream/downstream partnerships. Dynamic edges usually describes the event-based relationships that happens at arbitrary time points.  
    \item Hyperedges: Apart from pairwise edges, financial entities are also involved in set-based relations, which are expressed as hyperedges. Assets like stocks are often categorized based on their sectors, such as technology, real estate and healthcare. They are also segmented based on related concepts \cite{liu-zhang-2018-mining}. For example, the concept ``Metaverse'' is related to a set of stocks such as the Meta Platform, Nvidia, Unity, and Roblox.\footnote{\url{https://bit.ly/3OP5QRu}} Formally, a hyperedge $\mathcal{R} = \{n_i\} \subseteq V$ is a subset of $V$ involving a few entities. Graphs containing hyperedges are termed \textit{hypergraphs}. As a natural extension of graph neural networks, hypergraph neural networks are also being actively studied both theoretically and practically.

\end{enumerate}

\paragraph{Alternative data}
Alternative data refer to the multimodal data, such as text, image, and speech that convey predictive information. Quant models can refine their investment decisions by tapping into alternative data, which offers unconventional insights into diverse perspectives about the financial market. For example, the news about Elon Musk owning 9.2\% of Twitter makes the price of Twitter close up 27\%.\footnote{\url{https://bloom.bg/3bzOABA}} The success of deep learning in domains such as computer vision and natural language processing brings us tools to make decisions based on such comprehensive, multi-modal information. We can apply deep learning to extract useful events from news \cite{ritter2012open}, predict the sentiment of a post \cite{liu-etal-2018-learning-domain}, *mining the lead-lag relationship from knowledge graph \cite{Matsunaga_Suzumura_Takahashi_2019}* , and count the number of customers visiting Costco in a day \cite{collins2000system}.

\paragraph{Simulation data}

To improve quantitative models, high-quality data are essential. However, acquiring large-scale real-world financial data is challenging due to limitations in limited availability, privacy concerns, and high costs \cite{kannan2024review}. In this context, simulation data offer a viable solution. Synthetic data enable better incremental training, robustness assessment, and risk testing for alpha model development. The main generation methods fall into four categories:

\begin{enumerate}[wide]
    \item Rule-based:

Early market simulation studies heavily relied on rule-based approaches, where predefined rules dictated trading behavior under specific price conditions, often assuming trend following or mean reversion~\citep{raberto01,palmer94}, or simplistic resample methods. While these models offer interpretability and ease of implementation, their rigidity limits adaptability to real-world market dynamics, which involve non-linear relationships and evolving dependencies~\citep{kannan2024review}. 
\item Time Series-based:
Time series-based methods use historical data to simulate market dynamics. The advancement of deep learning enables the capture of complex financial patterns, including Variational Autoencoders (VAEs)~\citep{rezende2014stochastic, kingma2013auto}, Generative Adversarial Networks (GANs)~\citep{fu2019time, zhang2022data, goodfellow2014generative} and diffusion models~\citep{huang2024generative}. These models facilitate market trajectory prediction~\citep{coletta22, coletta23} and the simulation of market fluctuations and risks~\citep{cont2022tail, xia2024market, vuletic2024fin, wiese2020quant}. While effective in preserving statistical properties, they still struggle to model causal interactions between market participants.

\item Order Flow-based: With improvements in market microstructure research and computational efficiency, order flow-based methods are emerging. These models simulate order arrivals, executions, cancellations, and limit order book dynamics to capture price formation and volatility~\citep{chiarella02, chiarella09, li20}. They aim to uncover market dynamics by modeling order interactions, with examples like DeepLOB~\citep{zhang2019deeplob}, MarS~\citep{li2024mars}, and DiGA~\citep{DiGA}. Challenges include modeling complexity, high computational costs, and difficulty capturing long-term economic cycles beyond short-term fluctuations.

\item Multi-Agent-based:
Recent simulation methods also focused on multiple agents, modeling traders, investors, and dealers with distinct goals and decision strategies~\citep{zhang2024ai, lux99, abides20}. These agents interact with each other and the market, producing emergent phenomena that mirror real-world markets. While capturing market complexity, these approaches struggle with realism-efficiency trade-offs and parameter calibration for long-term accuracy~\citep{wang17}.

\end{enumerate}

\subsubsection{Features}

\label{sec:review_current-practice_factor-mining}

Financial data are intrinsically noisy and large-scale \cite{black_noise_1986, magdon-ismail_financial_1998}, making it difficult to extract meaningful information directly. Hence, we use the attributes derived from the original data, namely factors \cite{wiki_factor-investing, ross_arbitrage_1976, fama_cross-section_1992}, to describe the asset from different financial aspects such as value, size, momentum \footnote{\url{https://bit.ly/2HZHofh}} \cite{chabot_momentum_2014}, reversal \footnote{\url{https://bit.ly/3PxLkFG}}, volatility \footnote{\url{https://bit.ly/2QoH8Oq}}, etc. Mining factors integrate information, forming the foundation for subsequent quantitative models \citep{tulchinsky2019finding}. From the perspective of deep learning, the factor mining step corresponds to feature engineering \cite{wiki_feature-engineering} for financial data, the typical workflow consists of three procedures including feature construction, feature extraction, and feature selection \cite{liu_feature_1998}. 

Feature construction aims to enhance model performance by creating more informative representations, known as factors, through the strategic combination and manipulation of raw data. Financial factors can be categorized into two types: 1) Symbolic factors, expressed through symbolic equations or rules \citep{kakushadze2016101}, and 2) Machine learning factors, derived from machine learning models \citep{zhang2017stock}. 

Traditionally, symbolic factors have been explored by human researchers, relying heavily on expert knowledge and intensive labor. To broaden the scope of exploration and expedite the factor mining process, researchers are increasingly turning to algorithmic factor mining. This approach, essentially a task of symbol regression, can be addressed with regression techniques such as genetic programming algorithms \citep{zhang2020autoalpha, brabazon2020applications, ren2024alpha} and neural symbol regression \citep{yu2023generating, cui2021alphaevolve}.

While symbolic factors offer higher readability, their effectiveness is often limited by the predefined scope of operand and operator space, convergence efficiency and factor diversity. By contrast, machine learning factors boast a more flexible representation, owing to their high-dimensional parameter space. This flexibility allows for a more nuanced and adaptive approach to factor construction in financial modeling. Researchers typically employ feature extraction models, such as encoder-decoder architectures \citep{duan2022factorvae}, to align with predetermined learning objectives, like predicting future returns or volatility. In this process, either the latent representation or the prediction output of the model can be utilized as a signal factor \citep{xu2021hist, xu2021rest}. The effectiveness of machine learning factors stems from their robust expressive capabilities, which are attributed to high-dimensional parameters and non-linear functions, coupled with faster search efficiency directed by target gradient optimization. However, the primary challenge with these factors lies in their limited interpretability and readability, which poses difficulties in risk management.

The goal of feature selection is to select a subset of features for dimension reduction, avoiding overfitting and improving model performance. For quantitative investment, this step is usually intended for selecting the best-performing factors individually. Hence, one commonly used method for feature selection is the filtering method \cite{das_filters_2001, hall_correlation-based_1999, koller_toward_1996}, which scores each feature according to certain criteria and the best-performing ones are retained. As for the criteria, the correlation between the factor and the actual return \cite{zhang_information_2020} is often used to measure its predictive capability. The correlations between selected factors should be minimized to eliminate redundant information.

It should be noted that factors are not always necessary. Some approaches directly model from raw data inputs in an end-to-end fashion. With the advancements in deep learning techniques, these methods are increasingly favored by researchers, we will will elaborate on this in subsection \ref{para:end-goal}.

\subsection{Model Prediction}

In the field of prediction, researchers have extensively employed various deep learning methods, many of which have demonstrated practical effectiveness in real-world markets. The essence of deep learning modeling lies in two core aspects: the model architecture and the optimization objective. The model architecture is selected based on the inherent relationships between the targets and the available data, and it must possess the expressive power to capture these complex interdependencies. Meanwhile, the optimization objective determines how the quality of the model’s output is evaluated and guides the training process toward effective predictions. For the task of price prediction, it is crucial to address two fundamental questions:
\begin{enumerate}[wide]
\item How is future price information embedded in historical data, and how can we effectively model these relationships?
\item How to define the optimization objective so that the prediction signals best support subsequent trading strategies?
\end{enumerate}

\subsubsection{Modeling Data Interdependencies}

\figureDataDependencies
Financial data inherently exhibits temporal and spatial correlations. Temporal correlations (Fig \ref{fig:temporal_patterns}) reflect how data points are related over time, such as trends, momentum, or reversals in market dynamics. Spatial correlations are about the interconnectedness between entities, such as stock interdependencies, market sector influences, and upstream-downstream relationships in industry supply chains. Asset prices are driven by a complex interplay of these temporal and spatial factors. Accurately modeling these dynamics can lead to improved forecasting, as it allows predictive models to harness both historical trends and cross-sectional relationships among market entities. To achieve this, researchers have developed diverse models that capture these complex interactions. Depending on the nature of the relationships they aim to capture, prediction model architectures are typically categorized as temporal models, spatial models, or spatiotemporal models.

\paragraph{To capture temporal patterns}
Modeling temporal patterns is essential because it enables us to extract valuable information from historical data, detect trends, and forecast future price movements based on time-evolving signals. Therefore, to capture these time-dependent dynamics, researchers construct time-series inputs using historical price and volume data from assets and apply temporal modeling techniques that aggregate information across successive time steps. These models typically rely on the assumption of time translation invariance, meaning that the aggregation rules and parameters remain consistent across different time intervals. Typical examples of temporal blocks include convolutional neural networks \cite{oord_wavenet_2016,sim2019deep,hoseinzade2019cnnpred, chen2018stock}, RNNs \cite{zhang_stock_2017,nikou2019stock}, and transformers \cite{wen_transformers_2022}, along with their respective variations. Moreover, many studies employ hybrid combinations of these blocks—for instance, CNN-LSTM\cite{oncharoen2018deep} architectures—to capture temporal dependencies across different receptive fields, effectively modeling both local and sequential temporal correlations.
Specialized models are also devised for irregular time intervals. For example, \cite{zheng2017capturing} uses fine-grained feature-level time span information to decay the effect of previous timesteps for making use of irregular time intervals. \cite{jiao2023microstructure} partition the order sequence into discrete segments and perform temporal signal extraction on these segments to effectively model market microstructure.

\paragraph{To capture spatial patterns}
Capturing spatial patterns is crucial because it allows models to leverage the relationships and dependencies among different assets and their corresponding sectors. By understanding how different entities interact, models can improve predictions by considering the influence of correlated market movements and sector-wide trends. To capture these spatial relationships, researchers adopt spatial modeling techniques that can be divided into two primary approaches: implicit and explicit methods. \textbf{Implicit methods} typically use self-attention mechanisms that evaluate the entire set of entities simultaneously without relying on predefined graph structures \cite{feng_temporal_2019}. In contrast, \textbf{explicit methods} represent these relationships using sparse graph structures—either constructed directly \cite{li_modeling_2020,xu_hist_2022}, or inferred via graph structure learning methods \cite{zhu_deep_2021,xu_hist_2022}. Graph neural networks (GNNs) \cite{kipf_semi-supervised_2017,xu_hist_2022} are then applied on the graphs to discern meaningful patterns. GNNs are particularly effective at handling complex graph structures, including those that involve hyperedges \cite{sawhney_spatiotemporal_2020}, by facilitating message passing across nodes. Most graph neural networks function by facilitating message passing on the graph. In addition, some researchers \citeN{10447394,daiya2024diffstock} have explored the use of generative models, such as diffusion models, to generate dynamic asset graph structures that simulate the complex, time-varying relationships among different assets. 
Spatial methods typically assume spatial position invariance at different scales. For instance, self-attention maintains invariance to permutations in entities' positions, whereas GNNs ensure node and edge permutation invariance, allowing them to operate on various relational structures. Global and local methods in modeling have complementary strengths in their receptive fields. 
\citep{Kim_So_Jeong_Lee_Kim_Kang_2019} using Graph Attention Networks (GATs) for large-scale financial graph structures combines graph modeling of prior relationships with attention mechanisms. This dual approach helps focus on crucial nodes, diminishing the influence of complex background noise and enhancing the signal-to-noise ratio.

\paragraph{To capture spatiotemporal interactions}
Modeling spatiotemporal interactions is critical because it enables models to capture both the evolution of individual assets over time and the dynamic relationships between different market entities simultaneously. This comprehensive approach improves predictive accuracy by integrating insights from both dimensions. To capture these dual dynamics, researchers have adopted spatiotemporal modeling techniques that fuse spatial and temporal information. There are primarily two ways to achieve this integration: the decoupled and coupled approaches. 
The \textbf{decoupled approach} independently encodes spatial and temporal features; for example, a hypergraph encoder might first be applied to capture spatial relationships, followed by gated temporal convolution to model time-series dynamics \cite{hou_st-trader_2021}. MATCC\cite{MATCC} also designed a correlation module composed of multiple layers of attention and mixer junction submodules, each dedicated to modeling temporal correlations, inter-asset relationships, market trends, and other related factors. In contrast, the \textbf{coupled approach} integrates spatial and temporal information concurrently, thereby capturing the interactions between the two dimensions more directly and yielding improved representations of market behavior \cite{sawhney_spatiotemporal_2020,wang2021hierarchical}.

\subsubsection{Defining Prediction Objectives}
Once a model architecture is established, it is equally important to define an appropriate optimization objective that guides the model toward effective and accurate predictions. In quantitative investment models, defining learning objectives is particularly challenging because key components of the investment workflow—such as portfolio optimization and order execution—are typically non-differentiable. To address this, prior research has generally classified training objectives into two main categories: intermediary targets and direct end-goal optimization. Intermediary targets primarily assess future price movements of stocks, subsequently utilizing these outputs as predictive signals for portfolio construction through optimization algorithms. The alternative approach focuses directly on optimizing the aggregate performance of the final investment portfolio, thereby ensuring alignment of the output with the ultimate investment goal. Each approach offers its own benefits and trade-offs for aligning model outputs with real-world trading objectives.

\paragraph{Intermediary targets}
Opting for intermediary targets as the primary objective in model building offers simplicity by negating the need to consider complex elements like downstream risk control and the indifferential nature of order execution. However, this approach necessitates close coordination with downstream tasks, potentially leading to the accumulation of errors or a misalignment between the predicted targets and actual trading requirements.
In modeling price movements, two main approaches are employed: individual trend analysis and relative ranking. The individual trend method focuses solely on the future price trajectory of a single asset, typically using the asset's future returns as the label. This approach can be modeled either as a regression or a classification task. Regression tasks \cite{zhang_stock_2017}, leveraging mean squared error as the loss function, provide more precise fitting values for expected future returns, thereby being more conducive to downstream tasks. Classification \cite{sawhney_spatiotemporal_2020}, on the other hand, segments return labels into multiple categories, like price increases or decreases, commonly using cross-entropy as the loss function. Given the inherently low signal-to-noise ratio in financial data, numerous studies have explored ways to enhance signal quality through techniques like data sampling and label denoising. For instance, the LARA framework proposed in \cite{zeng2021trade} employs locality-aware attention to extract more informative samples from the data and uses RA-Labeling to adjust the labels of noisy samples on a per-trade basis during training, thereby improving the predictor's accuracy.
While these methods smooth out data noise, it often results in a larger gap with subsequent optimization tasks, necessitating additional signal transformation.
Relative ranking \cite{feng_temporal_2019}, in contrast, focuses on an asset's relative position within a cross-sectional framework. This method is particularly synergistic with certain portfolio strategies, such as long-short hedging strategies. Loss functions in relative ranking are classified into two types: local ranking, exemplified by pair-wise ranking methods\cite{du2024explainable} which evaluate pairs of assets to determine superior future performance, and global ranking, which involves inputting a group of assets and optimizing for the correlation between predictions and actual future returns, or other metrics pertinent to learning-to-rank tasks.

\paragraph{End-goal optimization}

\label{para:end-goal}
End-goal optimization in portfolio modeling directly addresses final portfolio positions \cite{liu2023deep}, focusing on complexities like inter-stock position control, risk management, and return stability from the start. This approach, while complex, effectively minimizes error accumulation common in multi-step models. Models employing end-goal optimization analyze a group of assets as a single sample, using spatio-temporal modeling approaches to account for both time-series trends and inter-asset relationships. 
Outputs typically include multi-period portfolio positions, optimized against performance metrics like return rates and the Sharpe ratio. Despite its complexity, this direct approach ensures alignment with portfolio management goals for comprehensive investment strategy. Despite the potential of this method, current research is limited by data availability and task complexity, leading to modest outcomes so far. However, with advancements in large-scale model technologies across various fields, this end-to-end modeling approach holds significant potential for future breakthroughs.

\subsection{Portfolio Optimization}

Portfolio optimization seeks optimal asset allocation to balance expected returns against volatility, transforming predictions of asset states into actual portfolio construction. This task, traditionally reliant on return and volatility predictions from statistical models, has been extensively researched within mathematical finance and operations research, as discussed in subsection \ref{sec:dl:oe:traditional}. With the advent of deep learning, the field has witnessed a transformative shift towards data-driven approaches, leveraging the vast availability of financial data and computational advancements. Deep learning has been applied in two main ways: enhancing existing optimization components (subsection \ref{sec:dl:oe:learning}) and pioneering end-to-end methodologies for direct allocation generation (subsection \ref{sec:dl:oe:e2e}). These advancements signify a new era in portfolio optimization, combining traditional insights with the capabilities of modern deep learning.

\subsubsection{Traditional Approaches}
\label{sec:dl:oe:traditional}
% Mean-variance approach and EGR approach

Portfolio optimization began with Markowitz's Modern Portfolio Theory, which formulates optimal portfolio generation as a quadratic programming challenge. Subsequent methodologies, such as those based on Kelly's criterion, aim to maximize cumulative returns over multiple periods. This subsection divides the problem into two scenarios: single-period portfolio selection, focusing on MPT \cite{lai_survey_2022}, and multi-period portfolio issues, concentrating on the latter \cite{li_online_2014}.

\paragraph{Single-period portfolio: Mean-variance Approach}
A single-period portfolio optimization problem can be simplified as generating a position for next holding period that can maximize the expected return while minimizing the potential risk. The expected return is usually represented as the mean of asset returns and risk is represented as the various of these returns, and such notation leads to its name: mean-variance approach, which is intended to balance return and risk.

The framework proposed by Markowitz is a very simple framework, and it cannot accommodate many practical considerations. Therefore, many follow-up works have been proposed to improve the original framework. Some important problems include the following. 
\begin{enumerate}
    \item Adding regularization terms into the optimization objective to improve the robustness of the portfolio or reach new goals \cite{carrasco2011optimal, corsaro2019adaptive}. For example, control of transaction costs can be realized by adding regularization terms to minimize the turnover rate. The position size can also be regularized by injecting sparsity regularization.
    \item Encourage diversity in portfolio allocation to reduce potential risk \cite{schmidt2019managing}.
    \item Better estimate the covariance matrix. As risk measure, the covariance matrix of assets is usually estimated from historical data. However, such estimation can lead to a severe problem that the observations (e.g. the number of trading days) used for estimation are insufficient to generate reliable covariance. Hence, various methods have been proposed to address this issue \cite{menchero2019portfolio, pafka2004estimated}, such as many estimators proposed in statistics, and the factor model that leverages dimensionaility reduction to explain asset returns using a small number of factors.
    \item Better risk measures. Covariance has several limitations: it is hard to estimate, and it is not always an ideal risk measure that generalizes to every scenario. Hence, other risk measures have also been proposed, including value at risk (VaR), conditional value at risk (CVaR).
\end{enumerate}

\paragraph{Multi-period portfolio: Online learning and stochastic control}

In the multi-period setting, the focus has changed to maximizing the cumulative return across multiple holding periods. Instead of generating one portfolio vector that has been extensively polished, now we need to generate a series of portfolio allocations whose cumulative returns across multiple periods are maximized. Relevant techniques include:
\begin{enumerate}
    \item Online learning with heuristics: the portfolio allocation for current period can be computed via optimization. The optimization objective can be formulated based on some basic trading ideas, such as momentum (follow-the-winner) and mean-reversion effects (follow-the-loser)\cite{li_online_2014}.
    \item Trend representation and pattern matching: naively following the basic ideas may not be effective in some cases. Hence more complicated trend representations are proposed to indicate some new trading ideas. Based on the predicted asset patterns and market distribution, new optimization objectives can be formulated.
\end{enumerate}

On the other hand, if we can model the dynamics of asset price, then the multi-period portfolio optimization problem can be regarded as an optimal control problem, where the cost function to minimize is the negative cumulative return. However, in practice the exact dynamics are usually impossible to be accurately modeled, so uncertainty is introduced to allow for better flexibility, and this problem now becomes one of stochastic control.

\subsubsection{Learning-based Portfolio Optimization}
\label{sec:dl:oe:learning}
Given that market dynamics and patterns are stochastic and difficult to predict, deep learning can be incorporated into the framework presented above to enhance certain modules, namely improving the dynamics, the estimator, and the solver.

\begin{enumerate}
    \item \cite{imajo_deep_2021} run Markowitz on specific returns (do neutralization on market risk factors via `spectral extraction`), and predict the distribution of the specific returns to compute their expected mean and variance. After getting the mean and variance, a Markowitz can be conducted.
    \item \cite{ma_portfolio_2021} use neural networks for the prediction of returns and risks and apply traditional portfolio optimization (Markowitz/Omega) to generate the final positions.
    \item \cite{wang_continuous-time_2019} studies continuous-time continuous-action and continuous-space portfolio optimization with finite horizon formalized as a stochastic control problem, and a Gaussian policy with time-decaying variance is derived. The Gaussian-based policy is proven to be better than adaptive control algorithms and DNN RL trained using DDPG. The asset price is modelled under GBM (Geometric Brownian Motion) and hence the dynamics are described by SDE. It is hence a stochastic control problem.
\end{enumerate}

\subsubsection{End-to-end Portfolio Generation}
\label{sec:dl:oe:e2e}

In addition to some works that use deep neural networks to combine parameter estimation and portfolio construction to achieve an "end-to-end" effect \cite{uysal2024end, anis2025end}, the reinforcement learning model has achieved better market performance in the field of end-to-end portfolio optimization itself.
 Since there is no universal, fixed and explicit label in portfolio optimization, it eliminates the need for nontrivial label construction, in order to achieve a more flexible risk-return balance. Without the need for complex model and parameter tuning, reinforcement learning models are trained on direct trading end-to-end feedbacks. In this case, designing a reasonable objective function becomes important. We now discuss several methods to design these functions.

\paragraph{Return-only approaches} Considering only portfolio returns is the most intuitive way of modeling. \cite{jiang_deep_2017} extract features from cross-sectional data to get the score of each asset, and then perform a normalization to get a position vector, the reward is then computed as the dot-product between this position vector and the return.

\paragraph{Risk-adjusted returns} \cite{wang_alphastock_2019} considers risks in investment and incorporates the Sharpe ratio as the reward function. The whole model is trained via policy gradient methods by propagating the Sharpe loss to different model parts.

\paragraph{Transaction costs} \cite{zhang_cost-sensitive_2022} considers the additional cost brought up by consecutive position adjustments, and added a regularization term computed as the l1 norm between the differences of the positions between 2 neighboring ticks.

\paragraph{Diversity} \cite{niu_metatrader_2022} considers different portfolio management styles may have different strengths in different markets. In this way a meta learning strategy is used to select from a number of different portfolio models trained with datasets generated by trading experts with different styles.

\subsection{Order Execution}
The order execution system implements the results from portfolio optimization, serving as a bridge between theoretical calculations and actual positions. It involves strategically placing and completing orders with the objective of minimizing total trading costs, considering factors including market conditions, asset liquidity and order impact. Research on order execution is usually conducted on high-frequency data, such as limit order book (LOB). The actions taken by participants in financial markets have become
increasingly based on quantitative analysis and algorithms rather than any human decision making process \cite{donnelly2022optimal}, the prevailing methods could be delineated into traditional optimal control models and reinforcement learning models.

\subsubsection{Traditional Approaches}

Assuming that the dynamics of the limit order book, including both its inherent dynamics and the impact of market orders, can be analytically represented as the problem of executing with minimum cost. This optimal control framework can be divided into discrete and continuous models, both of which aim to find a dynamic trading strategy that executes transactions over a fixed period, with an optimal utility function reflecting a combination of cost and volatility.

\paragraph{Discrete model}

The fundamental discrete trading model, proposed by the works of Bertimas \& Lo \cite{bertsimas_optimal_1998} and Almgren-Chriss \cite{almgren_optimal_2000}, assumes the price dynamics to be a discrete arithmetic Brownian Motion with fixed trading cost. They assumed that market impact is the only endogenous factor and the price volatility is exogenous. It is assumed to be the result of market forces that occur randomly and independently during trading, since market participants could detect and adjust their order placement. 

Taking the revenue volatility of different trading strategies into account, there is a trade-off between impact and variance. The impact of risk on optimal execution could be constructed by solving the minimization problem of the expectation of shortfall for a given level of variance. For each level of risk aversion, there will be a uniquely determined optimal execution strategy expressed as the efficient frontier of optimal execution.

\paragraph{Continuous model}

Continuous model further magnifies the time steps of the discrete model to infinity, solving the Almgren-Chriss problem with the dynamic asset price following specific distribution like Geometric Brownian Motion. The optimization of this stochastic control system can be transferred into a linear quadratic problem on a complete filtered probability space. \cite{2010Optimal} considers discrete trading under continuous GBM process, \cite{2012Optimal} provides numerical solutions under GBM using a mean-quadratic-variation objective function, \cite{2016Optimal} proves optimality under geometric price dynamics.

% \cite{2011A_115}, \cite{2011OPTIMAL_115}, \cite{dammann2022optimal}.

% \subsubsection{Better Price Dynamics}
The above model operates under the assumption of a zero - drift random - walk price process. To achieve better price dynamics, it explores three approaches for scenarios where the underlying assumption fails. First, the price process might exhibit drift, indicating a potential directional view of traders. Second, it could display serial correlation. Third, a future regime shift due to events such as earnings announcements is anticipated. Based on these assumptions, some studies generalize initial conditions using more realistic price dynamics. Almgren \cite{Almgren2003Optimal} proposes a non-linear stochastic price impact model since the previous linearity assumption \cite{almgren_optimal_2000} deviate a lot from factual accuracy. \cite{Almgren2012Optimal} solves the optimal strategy numerically under specific assumption of market liquidity and volatility. More realistic representation of price dynamics including temporary mean-reverting \cite{2021Optimal} are also introduced to refine stochastic price impact to find the optimal strategies. 

\subsubsection{Reinforcement Learning Framework}

Reinforcement learning (RL) is effective in optimizing order execution due to its adaptability in dynamic, sequential decision-making environments. It enables end-to-end adaptation to changing market conditions by learning from the rewards associated with executed actions without relying on preliminary oversimplified model assumptions. Task components of RL are highly suited to the order execution process, corresponding relationship is as follows:

\begin{itemize}[leftmargin=*]
    \item State: Market variables of execution process, such as remaining time and current position.
    \item Action: A decision set of trading process, usually consists of submit, modify or cancel orders.
    \item Reward: The transaction cost and impact with a market order.
\end{itemize}

The reinforcement learning agent structure was first proposed by \cite{2006Reinforcement}, which used Q-learning algorithms to train an optimized strategy for finding different price levels in given limit order book data. \cite{2014A} adapted the solution from the Almgren-Chriss model to a dynamic strategy considering the specific current market microstructure using Q-learning. \cite{lin_deep_2020} proposed a Deep Q-Network (DQN)  based RL algorithm for order splitting, and \cite{ning_double_2021} applied a zero-ending inventory constraint to a double-DQN method.

In terms of policy-based algorithms, \cite{daberius_deep_2019} was an early work that studied Proximal Policy Optimization (PPO) in order execution tasks, distinct from DQN. Numerical experiments were conducted in different environments with various dynamics. \cite{fang_universal_2021} proposed an actor-critic style policy distillation algorithm, and the model was trained on multiple assets to obtain a universal trading strategy. \cite{2020Policy} used deep policy gradient methods to optimize strategies based on linear quadratic regulator problems. \cite{lin_end--end_2020} presented a more end-to-end approach, using PPO with the limit order book and inventory as the state to directly output actions.

There are also model-based and multi-agent solutions. \cite{lin_agent-based_2021} established a multi-agent simulation system for order execution, where agents were trained and compared using various RL methods. An empirical study in \cite{lin_investigating_2021} demonstrated the generalizability of DRL-based order execution agents, showing that different RL agents can transfer quite well.

The efficacy of different reinforcement learning algorithms has been validated across authentic datasets from multiple markets. The heightened focus and expeditious advancements of reinforcement learning aims to enhance order execution decisions across diverse financial markets, particularly in scenarios where participants possess restricted information regarding the market dynamics.

\subsection{Current Limitations and Future Directions}

As the field of deep learning in alpha research continues to evolve, several emerging trends and areas of research promise to further revolutionize this domain. These advancements aim not only to enhance predictive accuracy and reliability of exeuction models but also to streamline the investment process, improve interpretability, and adapt to the dynamic nature of financial markets. Below, we outline four pivotal areas that encounter limitations but represent promising avenues for future exploration.

\paragraph{Automated Machine Learning}
The development and tuning of predictive models for financial markets are both labor-intensive and complex, exacerbated by the markets' dynamic nature which necessitates frequent model updates. Automated Machine Learning (AutoML) presents a promising solution to streamline this process, reducing the need for manual intervention. By automating model selection, feature engineering, and hyperparameter tuning, AutoML can significantly enhance efficiency and adaptability in financial modeling, ensuring models remain relevant amidst rapidly changing market conditions.

\paragraph{Explanability}
% While deep learning models have stronger power than traditional methods, they are usually poor in explainablity. This brings up problem in finance where risk management is of great importance. Hence it is important to develop task-specific explainable AI methods to better open-up the deep learning black-box of AI.
Despite their superior predictive capabilities, deep learning models often lack transparency, making them less preferable in domains where understanding decision-making processes is crucial, such as finance. The importance of risk management in finance cannot be overstated, necessitating models not just for their predictive performance but also for their ability to provide insights into their predictions. Developing task-specific, explainable AI methods is essential to demystify the "black box" of deep learning, offering clear insights into model decisions and fostering trust among stakeholders.

\paragraph{Knowledge-driven AI}
% Deep learning relies on abundant data to outperform traditional methods. However, in investmnet a great proportion of scenarios cannot provide that large amount of data, for example long-term value investment. In such data-lack scearnios knoweldge-driven AI is needed to serve as a strong prior, in order to preserve comparative performance of deep leaning methods.

Deep learning's reliance on extensive datasets poses challenges in investment scenarios characterized by sparse data, such as long-term value investing. In these contexts, knowledge-driven AI can offer a valuable complement to data-driven approaches, integrating domain expertise as a robust prior to compensate for the lack of large datasets. 

\paragraph{End-to-end modeling}
While the pipeline consists of multiple stages, in practice models for these stages are usually trained separately with varying goals. such inconsistency may hinder a coherent optimization direction. Hence it is also promising to explore a fully end-to-end modeling method that takes raw data and generate trades directly, while being trained directly from the final return. In this way the goals are aligned and may lead to better training results.

\section{Large language models for alpha research}
The rapid development of large language models has taken quantitative investment a significant step forward, from AI-powered to AI-automated, with vast prospects for future development. However, their application remains less mature compared to other deep learning methods, with limitations in certain sub-tasks. In this section, we categorize their roles into predictor and agent for alpha research, and explores other potential applications in subsection \ref{llm:limit}.

% \subsection{LLM predictor}
\subsection{LLM-Based Predictor}

\label{llm:pred}
The emergence of Large Language Models (LLMs) has redefined the role of textual data in quantitative finance, transforming how investors extract insights and generate predictive signals. Traditional models primarily relied on structured numerical data, but LLMs now enable a deeper understanding of financial narratives, capturing market sentiment, causal relationships, and latent trading factors with unprecedented accuracy. This chapter explores how LLMs have evolved from passive sentiment classifiers to active market predictors, spanning applications from embedding-based sentiment extraction to direct factor generation for systematic trading. By bridging quantitative signals with qualitative reasoning, LLMs are paving the way for a new era of AI-driven market prediction—one where natural language understanding becomes a core component of financial modeling.

\begin{figure*}[h!]
    \centering
    \resizebox{\textwidth}{!}{  
        \begin{forest}
    forked edges/.style={edge path={(\forestoption{parent anchor}) -- +(3pt,0) |- (\forestoption{child anchor})}},
            for tree={
                grow=east,
                reversed=true,
                anchor=base west,
                parent anchor=east,
                child anchor=west,
                base=left,
                font=\small,
                rectangle,
                draw=black,
                rounded corners,
                align=left,
                minimum width=4em,
                edge+={darkgray, line width=1pt},
                s sep=3pt,
                inner xsep=2pt,
                inner ysep=3pt,
                ver/.style={rotate=90, child anchor=north, parent anchor=south, anchor=center},
            },
            where level=1{text width=10em,font=\scriptsize,}{},
            where level=2{text width=10em,font=\scriptsize,}{},
            where level=3{text width=10em,font=\scriptsize,}{},
            where level=4{text width=10em,font=\scriptsize,}{},
            [
            {LLMs-based Quant Predictor}, ver
                [
                {LLMs for Sentiment Extraction}, text width=9em
                    [
                    {Foundations of Alternative Data in Quantitative Analysis}, text width=16em
                            [
                        {Financial Time Series \& Micro-Blogging Correlations\cite{ruiz2012correlating}}, text width=16em
                                ]
                            [
                        {Wisdom of Crowds\cite{chen2013wisdom}}, text width=7em
                                ]
                            [
                        {Foundational Models: SentiWordNet\cite{baccianella2010sentiwordnet}, ANEW\cite{bradley1999affective}, VADER\cite{Hutto2015vader}}, text width=18em
                                ]
                        ]
                    [
                        {Embedding-based Classifiers}, text width=8em
                            [
                                {FinBERT (Araci)\cite{araci2019finbert}}, text width=6em
                            ]
                            [
                                {FinBERT (Zhuang \textit{et al.})\cite{ijcai2020p622}}, text width=8em
                            ]
                            [
                                {FinLlama\cite{konstantinidis2024finllama}}, text width=4em
                            ]
                            [
                                {Foundational Models: BERT\cite{devlin2019bert}, RoBERTa\cite{liu2019roberta}, Llama 2\cite{touvron2023llama}}, text width=17em
                            ]
                        ]
                    [
                        {Sentiment Extraction thru Prompts}, text width=10em
                            [
                                {Return Predictability and Large Language Models\cite{lopezlira2023chatgpt}}, text width=15em
                            ]
                            [
                                {Unveiling the Potential of Sentiment\cite{zhang2023unveiling}}, text width=12em
                            ]
                        ]
                    [
                        {Relational Representations}, text width=8em
                            [
                                {Static: \cite{kaur2023refind, vamvourellis2024company, rajpoot2023gpt}}, text width=6em
                            ]
                            [
                                {Dynamic: \cite{ouyang2024modal, xu2025modeling, li2023findkg}}, text width=7em
                            ]
                        ]
                    [
                        {Foundational Models: GPT-3.5\cite{brown2020language}, GPT-4\cite{openai2023gpt4}}, text width=13em
                        ]
                    ]
            [
            {LLMs for Financial Time-series Forecast}, text width=12em  % 修正 text, text width=12em 的错误
                    [
                        {Temporal Data Meets LLMs\cite{yu2023temporal}}, text width=10em
                        ]
                    [
                        {RiskLabs\cite{cao2024risklabs}}, text width=4em
                        ]
                    [
                        {Foundational Models: TIME-LLM\cite{jin2024timellm}, S2IP-LLM\cite{pan2024s2ip}},text width=15em
                        ]
                ]
            ]
        \end{forest}  
    }  
    \caption{LLM as predictors: an overview}
    \label{financial_architectures}
\end{figure*}
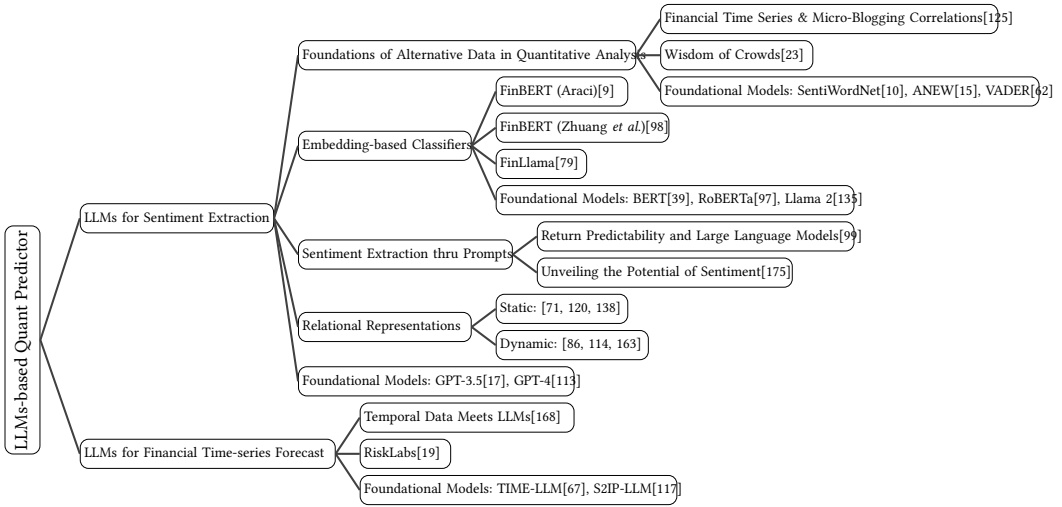

\subsubsection{LLMs for Sentiment Extraction}

\paragraph{Foundations of Alternative Data Analysis in Financial Markets}
Traditional quantitative models predominantly rely on structured market data to generate trading signals. However, the rise of alternative data—particularly from digital financial activities and online discussions—has introduced new dimensions to market analysis. As investor sentiment and information dissemination accelerate through news articles, analytical reports, and social media, these textual data sources have become increasingly influential in shaping market movements.

The predictive potential of textual data in financial markets was recognized well before the advent of Large Language Models (LLMs) \cite{chen2013wisdom, ruiz2012correlating}. Early approaches, however, lacked the computational sophistication needed to process large-scale textual information effectively. Initial methods relied on rudimentary techniques such as word frequency analysis and basic sentiment scoring \cite{schumaker2009textual}. For instance, \cite{gidofalvi2001using} applied a Naïve Bayesian classifier to map financial news articles to stock movement labels based on predefined time intervals. While these early models demonstrated the feasibility of extracting signals from financial text, their effectiveness was constrained by simplistic feature representations and limited contextual understanding. Prior to LLMs, sentiment analysis tools such as SentiWordNet \cite{baccianella2010sentiwordnet}, ANEW \cite{bradley1999affective}, and VADER \cite{Hutto2015vader} were widely used, but they struggled with contextual ambiguity and lacked deep semantic comprehension.

As deep learning methodologies advanced, researchers sought to enhance sentiment analysis pipelines by integrating machine learning and neural networks. A notable example is the ensemble deep learning model proposed by \cite{li2022ensemble}, which combined sentiment scores from VADER with a hybrid recurrent neural network (RNN) approach. Their blending ensemble model improved stock movement prediction accuracy by leveraging both textual news sentiment and stock price time-series data. A key contribution of their work was the exploration of historical news window sizes, reinforcing the notion that financial news exhibits long-memory effects, a characteristic later observed in LLM-driven sentiment models. These advancements paved the way for more context-aware and dynamic sentiment classification techniques, culminating in the emergence of LLM-powered financial text analysis.

Beyond sentiment scoring, recent work has expanded the role of causal relationships in financial forecasting. One significant advancement is the Causality-Guided Multi-Memory Interaction Network (CMIN)\cite{luo2023cmin}, which integrates financial text sentiment with causality-enhanced stock correlations to improve prediction accuracy. CMIN introduces a causal attention mechanism based on transfer entropy, ensuring that stock dependencies are not merely statistical artifacts but reflect directional information flows. The model’s multi-memory interaction framework allows textual and market correlation features to reinforce each other dynamically, bridging a key gap between traditional sentiment models and structured financial indicators. By demonstrating that textual sentiment alone is insufficient without context-aware stock relationships, CMIN represents a broader evolution in alternative data analysis, moving toward more interpretable and multi-modal forecasting techniques.

The emergence of LLMs has fundamentally reshaped this landscape, enabling more nuanced and context-aware textual analysis. By leveraging deep learning architectures, these models can capture complex semantic relationships, sentiment nuances, and implicit market signals embedded in financial text. This advancement has significantly enhanced the predictive capabilities of text-based trading strategies, marking a paradigm shift in alternative data analysis for financial markets.

One of the most direct and impactful contributions of Large Language Models (LLMs) to quantitative finance is their ability to extract sentiment from financial texts. Traditional quantitative models primarily rely on structured numerical data, which inherently lacks the contextual and qualitative aspects embedded in financial news, analyst reports, earnings calls, and social media discussions. The integration of LLMs into sentiment extraction can be broadly categorized into two key approaches: \textit{embedding-based classifiers} and \textit{LLM-powered sentiment generation}.

% \subsubsection{Embedding-Based Classifiers}

\paragraph{Embedding-Based LLM Classifiers}
One approach involves leveraging pre-trained word or sentence embeddings from models such as BERT\cite{devlin2019bert}, RoBERTa\cite{liu2019roberta}, or domain-specific variants like the two versions of FinBERT\cite{araci2019finbert,ijcai2020p622}. These embeddings serve as dense vector representations of financial texts, capturing semantic relationships between words and phrases. By training supervised classifiers (e.g., logistic regression, random forests, or neural networks) on these embeddings, researchers can map textual sentiment to predefined categories—such as \textit{bullish}, \textit{bearish}, or \textit{neutral}—with high precision. These classifiers benefit from efficiency and interpretability, making them ideal for large-scale applications in high-frequency trading and quantitative strategies. Further examples also include FinLlama\cite{konstantinidis2024finllama} a variant of Llama-2-7B\cite{touvron2023llama} fine-tuned on domain-specific financial text with a softmax activation layer for sentiment classification. In FinLlama, the fine-tuned model's sentiment classification is used directly as factor and back-tests were conducted to illustrate how more accurate sentiment classification can improve trading results. Another recent advancement in this direction is FININ\cite{wang2024finin}, which extends traditional embedding-based sentiment classification by modeling interactions among financial news articles rather than treating each piece of news independently. FININ constructs a Financial Interconnected News Influence Network, integrating multi-modal data (news text embeddings and market statistics) to assess the impact of both individual news items and their contextual interactions. Notably, FININ leverages pre-trained LLMs but does not fine-tune them for sentiment classification. Instead, it processes these embeddings through an influence quantifier, incorporating financial theories to assess market impact. By demonstrating that news interactions shape market sentiment, FININ represents an advancement in embedding-based models, bridging the gap between static embeddings and dynamic market-aware sentiment analysis.

% \subsubsection{LLM-Powered Sentiment Generation}

\paragraph{LLM-Powered Sentiment Classification}
An alternative, more dynamic approach involves directly prompting LLMs to generate sentiment classifications by incorporating financial news or earnings reports into structured queries. Unlike traditional methods that rely solely on pre-trained embeddings, this approach allows the LLM to contextually interpret sentiment within news text in relation to evolving market conditions. One of the most influential studies in this domain is \cite{lopezlira2023chatgpt}, which systematically evaluated ChatGPT’s ability to forecast stock price movements based on financial news sentiment. Their findings demonstrated that ChatGPT\cite{brown2020language }-generated sentiment signals exhibit predictive power, even after accounting for traditional asset pricing factors, with statistically significant results in backtesting. Additionally, they observed that ChatGPT’s sentiment assessments closely aligned with analyst consensus and market reactions, suggesting that LLMs can approximate human-like financial reasoning. To validate this, their experimental design involved prompting ChatGPT to classify financial news headlines as bullish, bearish, or neutral, constructing sentiment-based trading signals, and evaluating their predictive strength through an asset pricing regression framework and trading simulations.  

Building on this framework, \cite{zhang2023unveiling} applied a similar methodology to Chinese financial news, testing its effectiveness in a trading simulation of Chinese A-shares, further demonstrating the adaptability of LLM-driven sentiment analysis across different markets. Meanwhile, \cite{xie2023wallstreet} explored a hybrid approach, incorporating both structured quantitative data (e.g., stock prices and trading volume) and unstructured text from stock-related tweets to provide ChatGPT with a more comprehensive contextual understanding. Their work highlighted the potential for multimodal LLM-driven trading signals, combining sentiment cues with fundamental market data to enhance prediction accuracy.

% Besides financial news or social media, \cite{kim2024financial} have attempted at using LLMs to analyze financial statement and predict a company's future earnings, they further proceeded to construct porfolios adjustment strategies based on such predictions and showed that it yielded high returns. 
\paragraph{LLM-Based Relational Representation}

The integration of alternative data, such as news information, into relational frameworks (especially graph as Figure \ref{fig:relational_data}) has been shown to boost price prediction accuracy. Previously, graph relation construction was primarily based on real-world relationships, manual labeling, price correlations, or clustering algorithms. In contrast, LLMs' inherent comprehension and reasoning capabilities offer new perspectives for financial text understanding and market relationship modeling. 

In recent studies, \cite{kaur2023refind, vamvourellis2024company, rajpoot2023gpt} proposed the construction of static financial relations and validated the feasibility of extracting information from financial texts. In practice, more dynamic models with temporal variations are more widely adopted. \cite{ouyang2024modal, xu2025modeling} leverages information from video-audio and news sources respectively, establishing semantic relations. Based on this, predictions for market variables such as price, volatility and volume were made using a graph network approach. \cite{li2023findkg} propose FinDKG to extract global financial dynamic knowledge graph for risk tracking and investing with a better economic trend understanding.

\subsubsection{LLMs for Direct Time-Series Forecasting}

The integration of Large Language Models (LLMs) into time-series forecasting has introduced a paradigm shift in predictive modeling, leveraging their ability to process multi-modal data, capture complex dependencies, and provide human-readable explanations. Traditional deep learning architectures have demonstrated varying degrees of success in forecasting time-series movements. However, these models often struggle with cross-sequence reasoning, multi-source data fusion, and explainability, limiting their interpretability and robustness in dynamic environments. Recent studies have explored how LLMs can address these challenges, enabling more context-aware, interpretable, and adaptive forecasting frameworks.

A foundational step in this direction is the demonstration that LLMs can perform time-series forecasting in a zero-shot manner, requiring no explicit training on numerical sequences. \cite{gruver2023zeroshot} shows that LLMs, when prompted effectively, can match or exceed purpose-built forecasting models, highlighting their ability to generalize patterns and model complex temporal dependencies without additional fine-tuning. Extending this idea, \cite{jin2024timellm} introduces a reprogramming framework that aligns time-series data with the natural language capabilities of LLMs, enabling improved adaptability and performance across diverse forecasting tasks while keeping the backbone model intact. Further advancing this paradigm, \cite{pan2024s2ip} proposes S2IP-LLM, which aligns the semantic space of LLMs with time-series embeddings, enabling a contextualized prompt learning mechanism. These studies establish LLMs as generalizable time-series forecasters, setting the stage for their application in more specialized financial forecasting contexts.

Building on these foundational works, recent efforts have extended LLM applications to financial time-series forecasting, where market dynamics demand both predictive accuracy and interpretability. One of the pioneering studies in this domain, \cite{yu2023temporal} explores how LLMs can be applied to financial time-series forecasting with an emphasis on explainability. The study focuses on NASDAQ-100 stock prediction and proposes a methodology that integrates zero-shot and few-shot inference using GPT-4 to predict price movements, instruction-based fine-tuning of Open LLaMA, demonstrating that fine-tuning a public LLM can yield competitive results, and multi-modal data integration, incorporating stock price time-series data, financial news, and company metadata. A key finding is that GPT-4's reasoning capabilities, particularly when prompted with Chain-of-Thought (CoT) prompting, significantly improve predictive accuracy over traditional models like ARMA-GARCH and gradient boosting trees. The study also underscores the ability of LLMs to generate human-readable justifications for their forecasts, making them more transparent and interpretable compared to black-box machine learning models.

Beyond standard asset price prediction, the RiskLabs framework\cite{cao2024risklabs} extends LLM applications to financial risk prediction, integrating multi-modal financial data to estimate market volatility and Value-at-Risk (VaR). This model processes earnings conference calls (ECCs), analyzing both textual transcripts and vocal features (e.g., tone, sentiment), market-related time-series data, capturing price movements over multiple time horizons, and contextual news data, aligning financial reports and media sentiment with earnings announcements. RiskLabs employs an LLM-powered multi-task learning approach, where different modules extract, encode, and fuse insights from structured (numerical time-series) and unstructured (text/audio) data sources. Experimental results demonstrate that the RiskLabs model outperforms traditional statistical and machine learning models in volatility forecasting, highlighting the potential of LLM-driven multi-modal forecasting.

The application of LLMs to financial time-series forecasting represents a significant leap forward in market prediction and risk assessment. By integrating multi-modal data sources, fine-tuning for financial reasoning, and leveraging structured inference techniques, LLMs enhance both accuracy and interpretability in predictive finance. Future research should continue to explore hybrid approaches, combining LLM-powered reasoning with domain-specific econometric models, to further refine adaptive and explainable financial forecasting methodologies.

\subsection{LLM-Based Quant Agent}
\label{llm:agent}

% \subsubsection{Development of Financial LLMs and Agents}
The rapid development of large language models (LLMs) has catalyzed a paradigm shift in financial AI systems. Domain-specific financial LLMs like BloombergGPT \cite{wu2023bloomberggpt}, FinGPT \cite{yang2023fingpt}, and PIXIU \cite{xie2023pixiu} in English markets, along with CFGPT \cite{li2023cfgpt}, \cite{zhang2023xuanyuan}, \cite{wu2024golden} and DISC-FinLLM \cite{chen2023disc} in Chinese contexts, have demonstrated superior reasoning capabilities through pre-training on massive financial corpora and instruction-tuning on specialized tasks. These models have enabled the creation of standardized benchmarks for evaluating financial reasoning, sentiment analysis, and decision-making capabilities.

However, standalone LLMs face inherent limitations in real-time financial applications due to temporal latency and numerical processing constraints. This challenge has spurred the emergence of LLM-powered quant agents - AI systems that combine linguistic reasoning with tool invocation capabilities to process dynamic market data. These agents operate as autonomous entities capable of environmental perception, multi-step decision-making, and action execution through API integrations.

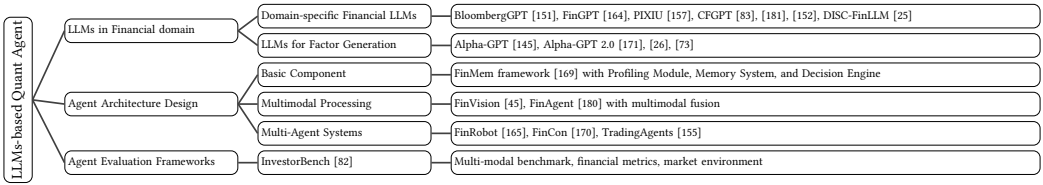
\begin{figure*}[h!]
    \centering
    \resizebox{\textwidth}{!}{
        \begin{forest}
    forked edges/.style={edge path={(\forestoption{parent anchor}) -- +(3pt,0) |- (\forestoption{child anchor})}},
            for tree={
                grow=east,
                reversed=true,
                anchor=base west,
                parent anchor=east,
                child anchor=west,
                base=left,
                font=\small,
                rectangle,
                draw=black,
                rounded corners,
                align=left,
                minimum width=4em,
                edge+={darkgray, line width=1pt},
                s sep=3pt,
                inner xsep=2pt,
                inner ysep=3pt,
                ver/.style={rotate=90, child anchor=north, parent anchor=south, anchor=center},
            },
            where level=1{text width=10em,font=\scriptsize,}{},
            where level=2{text width=10em,font=\scriptsize,}{},
            where level=3{text width=10em,font=\scriptsize,}{},
            where level=4{text width=10em,font=\scriptsize,}{},
            [
            {LLMs-based Quant Agent}, ver
                [
                {LLMs in Financial domain}
                    [
                        {Domain-specific Financial LLMs}
                            [
                                {BloombergGPT \cite{wu2023bloomberggpt}, FinGPT \cite{yang2023fingpt}, PIXIU \cite{xie2023pixiu}, CFGPT \cite{li2023cfgpt}, \cite{zhang2023xuanyuan}, \cite{wu2024golden}, DISC-FinLLM \cite{chen2023disc}}, text width=35em
                            ]
                        ]
                    [
                        {LLMs for Factor Generation}
                                [
                                    {Alpha-GPT \cite{wang2023alphagpt}, Alpha-GPT 2.0 \cite{yuan2024alpha}, \cite{cheng2024gpt}, \cite{kim2024financial}}, text width=35em
                                ]
                    ]
                    ]
                [
                {Agent Architecture Design}
                    [
                        {Basic Component}
                            [
                                {FinMem framework \cite{yu2024finmem} with Profiling Module, Memory System, and Decision Engine}, text width=35em
                            ]
                        ]
                    [
                        {Multimodal Processing}
                            [
                                {FinVision \cite{fatemi2024finvision}, FinAgent \cite{zhang2024FinAgent} with multimodal fusion}, text width=35em
                            ]
                        ]
                    [
                        {Multi-Agent Systems}
                            [
                                {FinRobot \cite{yang2024finrobot}, FinCon \cite{yu2025fincon}, TradingAgents \cite{xiao2024tradingagents}}, text width=35em
                            ]
                        ]
                    ]
            [
            {Agent Evaluation Frameworks}
                    [
                        {InvestorBench \cite{li2024investorbench}}
                            [
                                {Multi-modal benchmark, financial metrics, market environment}, text width=35em
                            ]
                        ]
                ]
            ]
        \end{forest}
    }
    \caption{Taxonomy of research in Financial LLMs and Agent Architectures.}
    \label{financial_architectures}
\end{figure*}

\subsubsection{LLMs for Direct Factor Generation}
While the previous sections focused on LLM-powered sentiment extraction from financial alternative data, the application of Large Language Models in quantitative investment extends beyond it. LLMs can also be leveraged to generate predictive factors directly as a factor agent, offering a novel approach to feature engineering in trading models. Instead of merely analyzing sentiment, these models can extract latent patterns from unstructured financial data, summarize qualitative insights into numerical indicators, and integrate LLM-generated signals into systematic trading strategies.
One of the most notable advancements in this area is Alpha-GPT \cite{wang2023alphagpt}, which introduces a human-AI interactive framework for factor mining in quantitative investment. Alpha-GPT leverages LLMs to assist researchers and traders in discovering novel alpha factors by engaging in an iterative dialogue, where the model proposes factor ideas, refines them based on human feedback, and generates executable code for implementation. This interactive workflow enables a more dynamic and adaptive approach to factor discovery, allowing domain experts to guide the AI’s reasoning while benefiting from its vast knowledge and pattern recognition capabilities. By incorporating domain knowledge and real-time market context, Alpha-GPT enhances the traditional factor discovery process, reducing reliance on purely statistical factor mining techniques. In subsequent versions, Alpha-GPT 2.0 emphasizes the iterative and interactive factor analysis process between humans and AI \cite{yuan2024alpha}. Leveraging LLMs, it automates the entire pipeline from alpha mining to modeling and analysis.
Building on this theme of LLM-driven factor generation, \cite{cheng2024gpt} explores how ChatGPT and GPT-4\cite{openai2023gpt4}can be positioned as a surrogate financial analyst to generate novel stock return factors. Instead of providing direct access to financial data, ChatGPT is only informed of the data structure and schema, ensuring that the generated factors are derived purely from its knowledge base rather than existing factor models. Through prompt engineering, the model is tasked with conceptualizing innovative stock return factors based on fundamental trading attributes. ChatGPT then outputs Python code to compute these factors, which researchers validate for originality and novelty before backtesting. This approach highlights LLMs’ capacity for creative financial factor discovery without relying on predefined quantitative models. Some of the other works in this realm also include \cite{kim2024financial} have attempted at using LLMs to analyze financial statement and predict a company's future earnings, they further proceeded to construct porfolios adjustment strategies based on such predictions and showed that it yielded high returns. 

\subsubsection{Architecture of LLM-based Quant Agents}
\begin{figure}[htbp]
\centering
\includegraphics[width=0.95\linewidth]{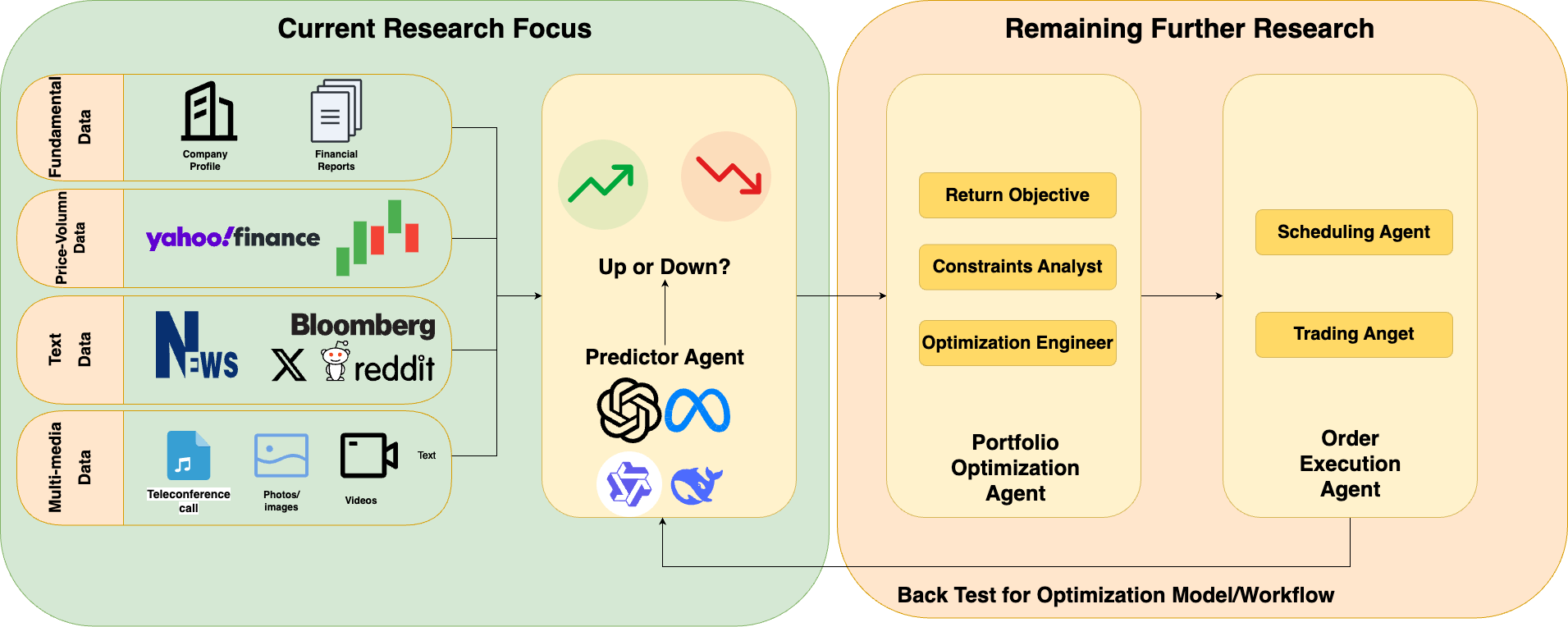}
\caption{Architecture overview of LLM-based quant agents.It has three parts: using data to predict price trends, optimizing asset choices, and making trades.}
\label{fig:agent_architecture}
\end{figure}

As shown in Figure.\ref{fig:agent_architecture}, a LLM-based Financial Agent integrates the three stages of financial decision-making into a cohesive process. The left part of Figure.\ref{fig:agent_architecture} details the data types used: fundamental (company profiles, financial reports), price-volume data collected from the data platform such as Yahoo Finance, text data(such as news, Bloomberg, Reddit and social messages from X.com), and multimedia (teleconferences, images, videos). These data feed into a Predictor Agent, which uses large language models to forecast stock trends (up or down).The right part of Figure.\ref{fig:agent_architecture} shows the subsequent stages. The Portfolio Optimization Agent uses stock predictions to allocate funds optimally, considering return objectives, constraints, and risk control. Finally, the Order Execution Agent implements these allocations by executing market orders efficiently, minimizing losses from market impacts. A feedback loop at the bottom indicates back-testing for refining the optimization model and workflow. This system integrates data analysis, prediction, optimization, and execution for robust, data-driven investment strategies. 

In the model prediction phase, the agent leverages large language models to analyze diverse data sources, enhancing prediction accuracy through LLMs and advanced deep learning techniques. Currently, the majority of scholarly investigations have focused on this particular aspect\cite{yu2024finmem,zhang2024FinAgent,yang2024finrobot}. 
These predictions then feed into the portfolio optimization stage, where the agent uses optimization algorithms to allocate assets, balancing risk and return based on investor preferences. Finally, in the order execution phase, the agent employs smart order routing and execution strategies to minimize market impact and maximize trade efficiency. However, both of these components still necessitate further in-depth research and exploration\cite{xiao2024tradingagents}. Throughout these stages, the LLM-based Financial Agent ensures data-driven, efficient, and adaptive decision-making, offering a significant edge in dynamic financial markets.

\subsubsection{The Expansion of LLM-based Agents}
The evolution of financial agent architectures has demonstrated remarkable progress through successive innovations and expansions in cognitive modeling and system integration. The foundational FinMem framework \cite{yu2024finmem} pioneered a basic architecture that established three core components now considered essential in modern systems. In the FINMEM framework, the design is structured around three core modules: The profiling module customizes the agent's characteristics by defining its professional background and investment risk inclinations, allowing it to adapt to market dynamics through dynamic risk settings. The memory module emulates human cognitive processes to handle hierarchical financial information, incorporating working memory for temporary storage and operations, and layered long-term memory for managing information with varying time sensitivities. This structure enables the agent to prioritize and retrieve critical memory events effectively. The decision-making module integrates these outputs to support informed investment decisions, prioritizing key memory events and aligning with market conditions for high-quality trading outcomes. Experimental results demonstrate that FINMEM significantly outperforms other trading agents, achieving superior cumulative returns and Sharpe ratios, particularly in handling complex financial data and adapting to market conditions.

Subsequent expansions addressed the growing complexity of financial data ecosystems through multimodal processing architectures. The FinAgent \cite{zhang2024FinAgent} framework integrates multimodal data, including numerical, textual, and visual information, to analyze financial market dynamics and historical trading patterns. It employs a dual-level reflection module to adapt to market changes and enhance decision-making by learning from historical data. The agent utilizes large language models (LLMs) to process market intelligence and incorporates tool-augmented strategies to guide trading decisions. Experiments on six financial datasets, including stocks and cryptocurrencies, demonstrate that FinAgent significantly outperforms state-of-the-art baselines, showcasing its effectiveness in handling diverse financial assets and its ability to generalize across various market conditions.
FinVision \cite{fatemi2024finvision} is a multi-modal multi-agent framework designed for stock market prediction. The framework integrates specialized Large Language Model (LLM) agents to process various financial data types, including textual news summaries, technical analysis of candlestick charts, and reflection on historical trading signals. The experimental results demonstrate that FinVision outperforms traditional buy-and-hold strategies and reinforcement learning models in terms of annualized returns and risk-adjusted metrics, such as the Sharpe Ratio. While it falls short of the benchmark set by the FinAgent model, FinVision achieves competitive performance with a significantly shorter training period, highlighting its potential for adaptability and robustness in dynamic market conditions. The framework's ability to manage risk effectively while optimizing returns underscores its value in complex financial environments.

The architectural frontier has recently shifted towards collaborative multi-agent systems that replicate institutional trading desk dynamics. FinRobot \cite{yang2024finrobot} is an open-source AI agent platform designed for financial applications, leveraging large language models (LLMs) to enhance financial analysis and decision-making. The platform is structured into four layers: Financial AI Agents, Financial LLM Algorithms, LLMOps and DataOps, and Multi-source LLM Foundation Models. The Financial AI Agents layer uses Chain-of-Thought (CoT) prompting to break down complex financial problems into logical sequences, while the Financial LLM Algorithms layer dynamically configures model application strategies. The LLMOps and DataOps layer ensures accurate models through training and fine-tuning techniques, utilizing task-relevant data. The platform's Smart Scheduler mechanism integrates various LLMs, selecting the most suitable models for specific tasks. Experimental results demonstrate the platform's effectiveness in market forecasting and document analysis, showing its ability to provide comprehensive insights and actionable recommendations for financial professionals. 
FINCON \cite{yu2025fincon} introduces a novel Large Language Model (LLM)-based multi-agent framework for financial decision-making, focusing on single-stock trading and portfolio management. The framework, inspired by real-world investment firm structures, employs a manager-analyst hierarchy to facilitate synchronized collaboration and reduce communication costs. It includes a dual-level risk-control component that monitors market risk and updates investment beliefs through self-critique. Experiments demonstrate that FINCON outperforms state-of-the-art LLM-based and DRL-based agents in terms of Cumulative Return (CR) and Sharpe Ratio (SR), while achieving one of the lowest Maximum Drawdown (MDD) values. Specifically, FINCON achieves a CR of 82.871\% and a Sharpe Ratio of 1.972\% for single-stock trading, and a CR of 113.836\% and a Sharpe Ratio of 3.269\% for portfolio management, showcasing its robustness and effectiveness in managing financial risks and enhancing trading performance.
The TradingAgents \cite{xiao2024tradingagents} framework introduces a novel multi-agent system for financial trading, leveraging Large Language Models (LLMs) to simulate the collaborative dynamics of a trading firm. The framework features specialized agents, including fundamental analysts, sentiment analysts, technical analysts, and traders, who collectively analyze market data and make informed trading decisions. Through structured communication protocols, agents engage in debates and discussions to reach balanced recommendations. Experimental results demonstrate that TradingAgents significantly outperforms baseline models in terms of cumulative returns, Sharpe ratio, and maximum drawdown, highlighting its ability to capture high returns while managing risk effectively. The framework's adaptability and explainability, facilitated by natural language-based operations, provide a distinct advantage over traditional trading strategies.

% \subsubsection{Evaluation Frameworks and Performance Benchmarks}
The maturation of financial agent research has driven systematic development of evaluation methodologies to address the domain's unique challenges. InvestorBench \cite{li2024investorbench} introduces a comprehensive benchmark for evaluating large language model (LLM)-based agents in financial decision-making tasks. The benchmark addresses the lack of a versatile framework and standardized datasets by providing a suite of tasks applicable to various financial products, including stocks, cryptocurrencies, and exchange-traded funds (ETFs). The method involves a structured LLM agent framework with modules for perception, profile, memory, and action, which processes and integrates market data, historical insights, and self-reflection to inform investment decisions. Experimental results show that proprietary LLMs generally outperform open-source and domain-specific models, particularly in volatile market conditions. Larger model sizes within the open-source category also demonstrate improved performance, highlighting the importance of model scale in financial decision-making. The benchmark provides a valuable platform for assessing and comparing the reasoning and decision-making capabilities of LLMs in complex financial scenarios.

\subsection{Current Limitations and Future Directions}
\label{subsec:limitations}
\label{llm:limit}

\subsubsection{LLM-Based Predictor}

While the application of LLMs in financial sentiment analysis and time-series forecasting has shown promise, several critical challenges remain. Addressing these limitations will require novel adaptations and hybrid approaches that blend LLMs with specialized financial models to improve interpretability, efficiency, and predictive accuracy.

One key challenge in sentiment-based financial predictions is the misalignment between linguistic sentiment and market sentiment. LLMs, trained primarily on textual corpora, interpret sentiment through linguistic cues rather than financial market reactions. Although textual sentiment can correlate with price movements, it does not always translate directly into market behavior due to factors like pre-existing investor expectations, macroeconomic conditions, or sector-wide trends. Moving forward, future research should focus on aligning LLM-derived sentiment with financial market dynamics, potentially through reinforcement learning, cross-modal attention mechanisms, or fine-tuning on financial-specific datasets. By integrating structured financial indicators—such as historical price action, volatility metrics, and earnings reports—into the sentiment extraction pipeline, LLMs could develop a more market-aware understanding of sentiment beyond purely linguistic interpretations.

Another key limitation stems from the representation of numerical time-series data within LLMs. Because LLMs process text-based tokens rather than continuous numerical sequences, financial data must often be reformatted to align with an LLM’s reasoning framework. However, time-series data has an inherently low signal-to-noise ratio, meaning that naive tokenization may introduce excessive noise, reducing predictive reliability. At this stage, the direct application of LLMs in time-series forecasting remains at the preliminary research, with limited adoption in the industry. How to leverage the reasoning capabilities of LLMs for forecasting remains one of the key research focuses. A promising direction is to develop more structured embeddings for numerical time-series data, possibly by integrating transformer-based numerical encoders or creating specialized financial LLM tokenization techniques that preserve key statistical properties of market signals.

Similarly, in quant prediction, one of the most pressing concerns is latency in real-time decision-making. While zero-shot prompting and in-context learning enable LLMs to generalize across different forecasting tasks, these methods often introduce computational overhead. In fast-moving financial markets, models must generate predictions within milliseconds to remain actionable. Future advancements should explore lightweight, fine-tuned LLM architectures or hybrid models that combine LLM reasoning with low-latency numerical models (such as Kalman filters or traditional time-series regressions) to enhance real-time responsiveness.

Moreover, financial time-series forecasting is fundamentally relational, as market movements are influenced not only by a single company’s historical prices but also by the behavior of its competitors, industry peers, and macroeconomic factors. Traditional econometric models explicitly encode these dependencies, but LLMs, when directly used for sequential forecasting, currently lack a clear mechanism to capture cross-company relationships. Future research should explore graph-based learning techniques, where LLMs incorporate relational embeddings that encode inter-stock dependencies. Alternatively, multi-modal LLM architectures that combine structured financial graphs with textual market news could enhance spatial dependency modeling and lead to more holistic market predictions.

In sum, while LLMs introduce powerful new capabilities for financial sentiment analysis and time-series forecasting, their current limitations highlight the need for domain-specific adaptations. Future research should focus on integrating market-aware sentiment modeling, optimizing real-time efficiency, developing structured numerical embeddings, and incorporating relational financial knowledge to further refine LLM-driven forecasting methodologies. By bridging the gap between text-driven reasoning and structured financial modeling, LLMs could evolve into more robust, interpretable, and adaptive tools for quantitative investment.

\subsubsection{LLM-Based Agent}

The integration of LLM-based agents into professional investment workflows has revealed significant gaps when evaluated against institutional-grade alpha generation pipelines (Section \ref{alpha_pipelne}). These limitations span the entire investment decision chain, from predictive analytics to execution dynamics, presenting both challenges and opportunities for future research.

The prediction paradigm represents a fundamental challenge for current LLM agents. Same as LLM-based predictors, despite remarkable natural language processing capabilities, LLMs exhibit notable deficiencies in quantitative reasoning tasks \cite{zhang2024FinAgent,fatemi2024finvision}. The numerical reasoning gap manifests in their limited ability to recognize precise price-volume patterns compared to specialized quant deep learning models, particularly in detecting subtle technical indicators and regime transitions. The first approach involves integrating existing deep learning models as tools into a specialized LLM quant-analyst agent, using the agent as the control center for generating and optimizing prediction models, thereby enhancing the accuracy of the agent's predictive and decision-making capabilities.
Furthermore, Emerging hybrid architectures that combine LLMs with quantile regression networks offer promising solutions, potentially enabling probabilistic forecasting while maintaining interpretability through natural language explanations of model outputs.

% This limitation is compounded by the fixed prediction windows employed by most existing systems, which fail to capture the multi-horizon nature of financial markets where short-term volatility patterns interact with long-term fundamental trends. 
% Furthermore, the causal inference capabilities of current LLM-based agents remain rudimentary, often conflating correlation with causation in complex market movements. 

Portfolio optimization frameworks in current LLM-based systems reveal significant deviations from institutional practice. Currently, trading agents \cite{yu2024finmem,zhang2024FinAgent} predominantly focus on the trading individual assets, while research on multi-asset trading and portfolio optimization remains notably underdeveloped. The risk constraint formulations employed by existing frameworks often rely on oversimplified models, failing to incorporate sophisticated risk measures such as Conditional Value at Risk (CVaR) . 
This limitation is exacerbated by the widespread neglect of transaction cost modeling, with most systems assuming frictionless markets. The diversification mechanics in current multi-agent systems lack explicit constraints, potentially leading to concentration risks that would be unacceptable in institutional portfolio optimization. 
The most straightforward solution is, since existing portfolio optimization techniques are relatively mature, to integrate them as tools into a specialized LLM optimization agent, enabling the agent to invoke these methods, analyze the results, execute decisions, evaluate and iterate autonomously, thereby enhancing the overall optimization performance of the agent. 
Neuro-symbolic approaches also present a promising direction for bridging this gap, potentially enabling the translation of LLM-generated market theses into mathematically rigorous optimization constraints while maintaining the flexibility of machine learning models.

The order execution stage exposes critical operational limitations in current agent designs. Most systems operate under the unrealistic assumption of perfect liquidity, ignoring the complex dynamics of limit order books and the substantial market impact of large orders \cite{yu2024finmem,yu2025fincon,zhang2024FinAgent}. This oversight is particularly problematic for liquid instruments where execution cost can significantly impact overall performance. The lack of real-time inference and adjustment mechanisms further compounds these issues, as agents fail to adapt their execution in time to changing market conditions. 
Integrating advanced order execution tools and market microstructure simulators with low-latency LLM reasoning modules could potentially enabling llm trading agents to develop sophisticated execution strategies that account for both market impact and opportunity cost.

The evolution of LLM-based quantitative agents must address these limitations through a holistic approach that considers the entire investment pipeline. Future research directions should focus on developing hybrid architectures that combine the strengths of LLMs in natural language processing and pattern recognition with the quantitative rigor of traditional financial models. This integration should span all stages of the investment process, from predictive analytics that incorporate both fundamental and technical factors, through portfolio construction frameworks that implement institution-grade risk management, to execution systems that account for real-world market frictions. Such advancements would bridge the gap between academic research and professional practice, potentially enabling the development of next-generation quantitative investment systems that combine the flexibility of machine learning with the robustness of traditional financial engineering.

% \subsubsection{Research Opportunities}
% To address these limitations, we propose three key research directions:
% \begin{enumerate}
%     \item \textbf{Neuro-Symbolic Architecture Design}: Combining LLMs' linguistic reasoning with:
%     \begin{itemize}
%         \item Numerical solvers for convex optimization
%         \item Stochastic control modules for order execution
%         \item Causal discovery engines for regime detection
%     \end{itemize}
%     \item \textbf{Temporal Graph Integration}: Developing time-aware knowledge graphs that connect:
%     \begin{itemize}
%         \item Corporate event timelines
%         \item Macroeconomic indicator cadence
%         \item Market microstructure patterns
%     \end{itemize}
%     \item \textbf{Multi-Objective Benchmarking}: Expanding InvestorBench \cite{li2024investorbench} to include:
%     \begin{itemize}
%         \item Risk-adjusted return metrics (e.g., Omega ratio)
%         \item Market impact simulations
%         \item Regulatory compliance checks
%     \end{itemize}
% \end{enumerate}

% Recent work by \cite{kim2024fusion} demonstrates the potential of these directions, achieving 39\% error reduction in price prediction through hybrid LLM-GARCH architectures while maintaining natural language explanations of volatility regimes. The field now stands at the threshold of creating truly holistic quant agents that can navigate the complete alpha pipeline from data ingestion to executed trades.

\section{Conclusion}

Quantitative investment, particularly alpha strategy, as a forefront technology in financial markets, is attracting increasing attention. In this survey, we provide an in-depth and comprehensive review of how deep learning and large language models (LLMs) are being studied and applied in this domain. We present a thorough coverage of alpha strategy research, including data, models, and each components of the overall pipeline. Building on this foundation, we examine the transformative impact and performance improvements brought by deep learning in various aspects. In addition, we explore how LLMs can be utilized most effectively as predictors and agents. Finally, we compare different stages of development and their applications in quantitative investment models, summarizing the current limitations, analyzing the prevailing challenges and intricacies, and discussing a series of future research perspectives.

% \end{spacing}{2.5}
% Bibliography
\newpage
\vskip 0.2in
\bibliographystyle{ACM-Reference-Format}
\bibliography{main}

\end{document}